\documentclass[submission, Phys]{SciPost}

\hypersetup{
    colorlinks,
    linkcolor={red!50!black},
    citecolor={blue!50!black},
    urlcolor={blue!80!black}
}
\usepackage{dsfont}
\usepackage{mathtools}
\usepackage{amsmath}
\usepackage[bitstream-charter]{mathdesign}
\DeclareSymbolFont{usualmathcal}{OMS}{cmsy}{m}{n}
\DeclareSymbolFontAlphabet{\mathcal}{usualmathcal}

\newcommand{\calB}{\mathcal{B}}
\newcommand{\calH}{\mathcal{H}}
\newcommand{\calE}{\mathcal{E}}
\newcommand{\frake}{\mathfrak{e}}
\newcommand{\frakE}{\mathfrak{E}}

\begin{document}

\begin{center}{\Large \textbf{
Finite-size corrections to the energy spectra of gapless one-dimensional systems in the presence of boundaries}\\
}\end{center}

\begin{center}
Yifan Liu\textsuperscript{1$\star$},
Haruki Shimizu\textsuperscript{1},
Atsushi Ueda\textsuperscript{2} and 
Masaki Oshikawa\textsuperscript{1,3,4}
\end{center}

\begin{center}
{\bf 1} Institute for Solid State Physics, University of Tokyo, Kashiwa 277-8581, Japan\\
{\bf 2} Department of Physics and Astronomy, University of Ghent, 9000 Ghent, Belgium\\
{\bf 3} Kavli Institute for the Physics and Mathematics of the Universe (WPI),
The University of Tokyo, Kashiwa, Chiba 277-8583, Japan
\\
{\bf 4} Trans-scale Quantum Science Institute, University of Tokyo, Bunkyo-ku, Tokyo 113-0033, Japan
\\
${}^\star$ {\small \sf yifan@g.ecc.u-tokyo.ac.jp}
\end{center}

\begin{center}
\today
\end{center}


\section*{Abstract}
{\bf
We present the finite-size scaling theory of one-dimensional quantum critical systems in the presence of boundaries.
While the finite-size spectrum in the conformal limit, namely of a conformal field theory 
with conformally invariant boundary conditions, is related to the dimensions of
boundary operators by Cardy, the actual spectra of lattice models
are affected by both bulk and
boundary perturbations and contain non-universal boundary energies.
We obtain a general expression of the finite-size energy levels
in the presence of bulk and boundary perturbations.
In particular, a generic boundary perturbation related to the energy-momentum tensor gives
rise to a renormalization of the effective system size.
We verify our field-theory formulation by comparing the results with the exact
solution of the critical transverse-field Ising chain and with accurate numerical
results on the critical three-state Potts chain obtained by Density-Matrix Renormalization Group.       
}

\vspace{10pt}
\noindent\rule{\textwidth}{1pt}
\tableofcontents\thispagestyle{fancy}
\noindent\rule{\textwidth}{1pt}
\vspace{10pt}

\section{Introduction}
\label{sec:intro}
Unraveling the complexities of strongly correlated quantum many-body systems represents a formidable yet critical goal in physics. Unlike free theories, these strongly correlated models often defy exact solvability, posing significant challenges to theoretical understanding. Consequently, we often rely on numerical simulations, focusing particularly on low-energy states that are pivotal for comprehending the essence of these strong correlations.

Simulating infinitely large systems, however, is an impractical endeavor, even though the true critical behavior of these systems becomes apparent only in the thermodynamic limit. As a result, finite-size scaling theory emerges as an indispensable tool for extrapolating results to the thermodynamic limit. Among the numerical methods available, the infinite density matrix renormalization group (iDMRG)~\cite{refs_on_DMRG,mcculloch2008infinite} stands out for its capability to simulate infinite systems with considerable accuracy, especially in finitely-correlated systems. Nonetheless, iDMRG encounters limitations near criticality due to the finite correlation length $\xi(\chi)$ associated with the finite bond dimension $\chi$. Significant numerical errors can arise if the model’s inherent correlation length exceeds $\xi(\chi)$.

To address this issue, finite-entanglement scaling,
in which $\xi(\chi)$ instead of the system size $L$ is used as a scaling variable has been proposed~\cite{PhysRevB.78.024410,PhysRevLett.102.255701,PhysRevB.86.075117,PhysRevB.91.035120,sherman2023universality,ueda2023finite,huang2023emergent}.
However, the detailed understanding of its scaling properties remains an open question.
Conversely, traditional DMRG effectively simulates finite systems, accurately capturing critical systems as long as $L<\xi(\chi)$,
and thus offers a route to obtain reliable insights into the thermodynamic limit by the finite-size scaling of data for the moderate system size $L< \xi(\chi)$ which are free from the
error due to the finite bond dimension $\chi$.

When the critical point is described by a conformal field theory (CFT) in $1+1$ dimensions, the finite-size scaling is particularly powerful,
thanks to the ingenious use of conformal mapping between the infinite system and a finite strip initiated by Cardy~\cite{cardy1984conformal,cardy1986operator}.
The energy spectrum of CFT on a finite strip with the periodic boundary conditions (PBC)
is related to the scaling dimension of bulk operators in the CFT and that with the open boundary conditions (OBC)
is to the scaling dimension of boundary operators in the boundary conformal field theory (BCFT)~\cite{cardy2004boundary}.

An actual finite-size lattice model, however, is not an exact realization of the CFT even at the critical point.
The effective field theory for such a system is given by the CFT with various irrelevant perturbations.
A precise analysis of the finite-size spectrum requires a consideration of such perturbations.
This is particularly important when there is a marginally irrelevant perturbation that causes notorious logarithmic corrections.
For the PBC, such a program has been successfully implemented.
Even with the exact diagonalization, which can handle very small systems, the finite-size scaling based on the CFT
has led to a very precise determination of the critical point~\cite{nomura1995correlation,OKAMOTO1992433,nomura1994critical,nomura1998symmetry,matsuo2006berezinskii}.
It has also been applied to larger system sizes using density-matrix renormalization group (DMRG)~\cite{PhysRevB.86.024403,PhysRevB.94.235155,PhysRevB.106.155106} and tensor-network renormalization (TNR) schemes~\cite{ueda2021resolving,ueda2023finite}.

However, such a systematic study of the finite-size spectrum beyond the exact BCFT is lacking for OBC and thus would be a powerful tool for the numerical study of boundary critical phenomena.
Meanwhile, in the presence of the boundaries, the analysis becomes more complicated due to the bulk and boundary perturbations.
Moreover, as we will discuss later in detail, there are additional difficulties due to the renormalization of the effective system size, which is also referred to as an extrapolation length in Refs.~\cite{cardyQuantumQuenchesCritical2016,ziffEffectiveBoundaryExtrapolation1996,PhysRevB.83.134425}.

In this paper, despite the inherent difficulties, we construct a systematic theory for the finite-size scaling of energy spectra in one-dimensional quantum systems with boundaries. Our approach incorporates the effects of bulk and boundary perturbations within the framework of BCFT. A critical aspect of our analysis is the use of the effective system size 
$L$, which surprisingly often differs from the number of lattice sites $N$. Nevertheless, it is the renormalized system size $L$ that yields accurate predictions of finite-size effects. To validate our theoretical framework, we compare our predictions with numerical simulations of the quantum transverse-field Ising (TF-Ising) and three-state Potts models.

We note that, although BCFT has been studied in various topics~\cite{doreyFiniteSizeEffects2000a,gaberdielConformalPerturbationTheory2009,fukusumiOpenSpinChain2021,konechnyRGBoundariesCardy2023},
a detailed systematic study of the finite-size spectrum in the general context is lacking.
For the TF-Ising model, there exist some articles that relate finite-size corrections to conformal perturbation theory instead of the boundary one \cite{izmailianBoundaryConditionsAmplitude2009,izmailianIsingModelMixed2009,izmailianFinitesizeCorrectionsIsing2010,izmailianUniversalAmplitudeRatios2012}.
Although they provide correct universal ratios, we will show that such results are only correct for specific operators. Working in the context of BCFT is indispensable for generic perturbation in other models.

This paper is structured as follows. In Section~\ref{sec:sum}, we summarize the main results of this paper and validate them through applications to the TF Ising and three-state Potts models, focusing on finite-size corrections under various conformal invariance boundary conditions. In Section~\ref {sec_perturbation}, we establish a framework for analyzing the finite-size effects of bulk and boundary perturbations on the energy spectrum of a CFT on an open segment.
To relate this Hamiltonian with the real lattice model, two key ingredients are highlighted in Section~\ref {sec_L}.
We conclude this paper by discussing the possible future directions based on this work in Section \ref{sec_dis}.

We leave some detailed derivations in the Appendix. Appendix~\ref{app:K-W} and \ref{app:dual_three} review the Kramers-Wanier and Kramers-Wanier-like duality in the TF-Ising and three-state Potts models, respectively. We derive the ground-state energy for the TF-Ising model with a small boundary magnetic field in Appendix~\ref{app: magnetic}. Operator product expansion (OPE) coefficients in the $M(6,5)$ minimal model used in this paper are provided in Appendix~\ref {app_OPE}, offering valuable insights for related research areas.
Our numerical calculations were performed using ITensor library~\cite{10.21468/SciPostPhysCodeb.4}.

\section{Summary of the main results and their applications}
\label{sec:sum}

Here, we present the main results of the paper.

\subsection{Boundary CFT spectrum with bulk and boundary perturbations}
\subsubsection{General aspect}
We are interested in studying the finite-size effects on the energy spectrum of lattice models. In the vicinity of criticality, this spectrum can be well approximated by the Hamiltonian of continuous field theory. In this context, the universal properties in the thermodynamic limit and finite-size effects are represented by the scale-invariant CFT Hamiltonian $H^{\text{CFT}}_{\alpha\beta}$ and its perturbations, respectively. To elaborate, we consider the spectrum of a CFT on a finite open segment with bulk and boundary perturbations. Specifically, we examine the Hamiltonian:
\begin{align}
    \calH^\text{eff}_{\alpha\beta}(L)=H^{\text{CFT}}_{\alpha\beta}
    +\sum_i g_i \int_0^{L} \mathrm{d}{v}\, \Phi_{i}(v)
    +\sum_j g^L_j \Psi_j^{\alpha\alpha}(0)+\sum_j g^R_j \Psi_j^{\beta\beta}(L)\ ,
\end{align}
where the system is defined on the open interval $L$ with the conformally invariant
boundary conditions $\alpha$ and $\beta$ at each end.
$\Phi_i$ is a bulk operator of CFT, and 
$\Psi_j^{\alpha\alpha}$ and $\Psi_j^{\beta\beta}$ are boundary operators of CFT living on $\alpha$ and $\beta$ boundaries, respectively.
\begin{align}
    \Psi_{j}^{\alpha \alpha}(x) \Psi_{j}^{\alpha \alpha}(y)&\sim\frac{1}{|y-x|^{2h_j}}+\cdots\ ,\quad y>x\ ,
\label{eq:boundaryOP_normalization}    
    \\
\Phi_{i}(x) \Phi_{i}(y)&\sim\frac{1}{(z_
    1-z_2)^{2h_i}(\bar{z}_
    1-\bar{z}_2)^{2\bar{h}_i}}+\cdots\ .
\label{eq:bulkOP_normalization}
\end{align}
\noindent These operators are normalized properly by the short-distance correlation functions
as bulk and boundary operators, respectively.
In the conformal limit ($g_j=g^L_j=g^R_j=0$), each energy eigenstate with the
given conformally invariant boundary conditions $\alpha$ and $\beta$ corresponds
to a boundary operator $\psi^{\alpha \beta}_n$.
When $\alpha \neq \beta$, it is a boundary-condition changing operator.
The ground-state energy generally has the bulk contribution determined by
the bulk energy density $\frake_0$ and the boundary energies $\frakE^{\alpha,\beta}$
as
\begin{align}
     \frake_0 L + \frakE^\alpha + \frakE^\beta\ ,
\end{align}
which is non-universal.
The universal part of the energy eigenvalue $\calE^{\alpha\beta}_n$ in the conformal limit,
after subtracting the non-universal bulk and boundary energies,
is determined by the boundary scaling dimension $h_n$
of the boundary operator $\psi^{\alpha \beta}_n$, as $(\pi/L)\left( h_n - \frac{c}{24} \right) $.

We then consider the perturbative expansion of the energy eigenvalue as
\begin{align}\label{eq:E_general}
    \calE^{\alpha \beta}_n &\sim
   \frac{\pi}{L} \left( h_n - \frac{c}{24} \right) 
    + \left(\delta \calE^{\alpha \beta}_n\right)^{(1)}_\text{boundary}
    + \left(\delta \calE^{\alpha \beta}_n\right)^{(1)}_\text{bulk}
    + \left(\delta \calE^{\alpha \beta}_n\right)^{(2)}_\text{boundary}
    + \cdots\ ,
\end{align}
up to the first order in the bulk perturbation $g_i$ and the second order
in the boundary perturbation $g^{L,R}_j$.
We note that the bulk perturbation $g_i$ is independent of the boundaries
and the same perturbation leads to finite-size correction of the energy spectrum
for the periodic boundary condition.
Therefore, we can determine the bulk perturbation $g_i$ from the energy spectrum
for the periodic boundary condition to evaluate its effect on the energy spectrum
on a system with boundaries.
This is useful for extracting the effect of boundary perturbations from the energy
spectrum, as we will demonstrate in Sec.~\ref{Sec_3-state_fixed}.

We find the perturbative corrections as follows\footnote{For most of the discussions in this paper, we assume that the bulk/boundary operators are primaries just for simplicity. One can easily obtain the results for descendants by applying Virasoro operators to the correlation functions.}:
\begin{align}\label{eq:bound_1st}
\left(\delta \calE^{\alpha \beta}_n\right)^{(1)}_\text{boundary} &=
   \sum_j g^L_j \left(\frac{\pi}{L}\right)^{h_j} C_{n j n}^{\alpha \alpha \beta}
   +
    \sum_j g^R_j \left(\frac{\pi}{L}\right)^{h_j} C_{n j n}^{\alpha \beta \beta}\ ,
\end{align}
where $h_j$ is the boundary scaling dimension of
$\Psi_j^{\alpha\alpha}$ and $C_{ijk}^{\alpha\beta\gamma}$ are the boundary OPE coefficients~\cite{lewellenSewingConstraintsConformal1992}:
\begin{align}
    \Psi^{\alpha\beta}_k(x_1)\Psi^{\beta\gamma}_l(x_2)=\sum_l C_{klm}^{\alpha\beta\gamma}(x_1-x_2)^{h_m-h_k-h_l}\Psi_m^{\alpha\gamma}(x_2),\quad x_1>x_2\ .
\end{align}
And:
\begin{align}
\left(\delta \calE^{\alpha \beta}_n\right)^{(1)}_\text{bulk} &=
\sum_i g_i\left(\frac{\pi}{L}\right)^{\Delta_i -1}\left\{\int_0^{\pi/2} \mathrm{d} \theta\, e^{i\Delta_i\theta}\sum_p {^\alpha B_{i}^p}C^{\beta\alpha\alpha}_{npn}F^{i\bar{i},nn}_{p}(1-\exp(2i\theta ))\right.\notag\\
     &\quad\quad\quad\quad\quad\quad\quad+\left. \int_{\pi/2}^{\pi} \mathrm{d} \theta\, e^{i\Delta_i\theta}\sum_p {^\beta B_{i}^p}C^{\beta\beta\alpha}_{npn}F^{i\bar{i},nn}_{p}(1-\exp(2i\theta ))\right\}\ ,
\end{align}
where $\Delta_i$ is the total bulk scaling dimension of $\Phi_{i}$
(in this paper, we denote the total scaling dimension, namely the sum of
the conformal dimensions of the holomorphic and antiholomorphic parts of the
bulk field, by $\Delta_i$),
$F$ is the conformal block \cite{difrancescoConformalFieldTheory1997} 
and $^\alpha B^j_{i}$ is the bulk-boundary OPE coefficient~\cite{cardyBulkBoundaryOperators1991,lewellenSewingConstraintsConformal1992}:
\begin{align}\label{bb-OPE}
    \Phi_{i}(z)=\sum_j {^\alpha B^j_{i}}|z-\bar{z}|^{h_j-h_i-\bar{h}_i}\Psi^{\alpha\alpha}_j(x)\ ,
\end{align}
for the boundary condition $\alpha$ imposed on the real axis. 
For the ground state $n=0$ with diagonal boundary condition $\alpha$, this can be explicitly evaluated as 
\begin{align}
\left(\delta \calE^{\alpha \alpha}_0 \right)^{(1)}_\text{bulk} &\sim
\sum_i g_i \left(\frac{\pi}{L}\right)^{\Delta_i-1} 
{}^{\alpha}B_{i}^{\mathds{1}}
\frac{\sqrt{\pi}\Gamma(1/2-\Delta_i/2)}{2^{\Delta_i}\Gamma(1-\Delta_i/2)}
+ \text{const.}\ ,
\label{eq:E0_bulk_1st}
\end{align}
where the (possibly UV divergent) constant can be absorbed
by the non-universal boundary energy.

The second-order correction to the ground-state energy for the identical boundary
condition $\alpha$ imposed at both ends is
\begin{align}
\left(\delta \calE^{\alpha \alpha}_0\right)^{(2)}_\text{boundary} &=
\sum_j \left(\delta \calE^{\alpha \alpha}_0\right)^{(2)}_{g^L_j, g^R_j }\ ,
\end{align}
where
\begin{align}\label{eq:bo_2nd}
\left(\delta \calE^{\alpha \alpha}_0\right)^{(2)}_{g^L_j, g^R_j } & \sim
- \left(\frac{\pi}{L}\right)^{2h_j-1} \left[
\left({g^L_j}^2+{g^R_j}^2\right)\frac{\Gamma(1/2-h_j)\Gamma(h_j)}{2^{2h_j}\sqrt{\pi}}
+\frac{2g^R_j g^L_j}{h_j}{}_2F_1(h_j;2h_j;1+h_j;-1)\right]\\\notag 
&\quad + \frac{\text{const.}}{a^{2h_j-1}} \left( {g^L_j}^2+{g^R_j}^2 \right) \ ,
\end{align}
for $h_j \neq \frac{1}{2}$, and
\begin{align}
\left(\delta \calE^{\alpha \alpha}_0\right)^{(2)}_{g^L_j,g^R_j} & \sim
 - \left[\left({g^L_j}^2+{g^R_j}^2\right)
   \left(\text{const.} +\log{\frac{L}{a}}\right)+ 2 g^R_j g^L_j \frac{\pi}{2} \right]\ ,
   \label{eq:boundary2nd_h0.5}
\end{align}
for $h_j=1/2$, where $a$ is the lattice spacing.
The (possibly UV divergent) constant terms proportional to ${g^L_j}^2+{g^R_j}^2$ can be
again absorbed as non-universal boundary energy corrections.

\subsubsection{Energy-momentum tensor perturbations}\label{sec:T_per}

Any CFT has the boundary operator of scaling dimension $2$, which can be
identified with the (holomorphic part of) the energy-momentum tensor $T$.
For the boundary perturbation $T$, the first-order correction to the energy eigenvalues
is 
\begin{align}
   \left(\delta \calE_n^{\alpha\beta}\right)_{T}= \left( \frac{\pi}{L} \right)^2(g^{\alpha}_T+g^{\beta}_T)\left(h_n-\frac{c}{24}\right),
\end{align}
where $g_T^{\alpha/\beta}$ is the coupling constant of $T$ associated with boundary condition $\alpha/\beta$.
However, this correction can be absorbed by renormalizing the system size
at least in the first order in $g_T$, since 
\begin{align}
    \calE^{\alpha\beta}_n(L)+\left(\delta \calE_n^{\alpha\beta}\right)_{T}(L)&=\left(\frac{\pi}{L}\right)\left[1+(g^\alpha_T+g^\beta_T)\left(\frac{\pi}{L}\right)\right]\left(h_n-\frac{c}{24}\right)\notag \\
    &\sim\frac{\pi}{L+(\delta L)_\alpha+\delta(L)_\beta}\left(h_n-\frac{c}{24}\right)\notag\\
    &=\calE^{\alpha\beta}_n(L+(\delta L)_\alpha+(\delta L)_\beta),
\end{align}
where we set the local shift of system size $(\delta L)_{\alpha/\beta}=-\pi g_T^{\alpha/\beta}$. Thus, as long as we use the appropriate effective length for a lattice model,
we do not have to consider this perturbation explicitly for all energy levels.

For a lattice model, this suggests that
effective system size $L$ is not generally identical to the naive
system size $Na$ of the lattice model, where $N$ is the number of sites and
$a$ is the lattice constant. Instead, we have
\begin{equation}\label{eq:def_L}
    L = \left[N + \left(\delta L\right)_L + \left(\delta L \right)_R\right]a\ ,
\end{equation}
where $\left(\delta L \right)_{L,R}$ are the non-universal corrections of order
of $O(1)$, determined by the boundary conditions on each end of the lattice model.

The CFT can also have a bulk perturbation:
\begin{equation}
    g_{T} \int \mathrm{d}y \; \left[ T(w) + \bar{T}(\bar{w}) \right] \ .
\end{equation}
However, this term is proportional to the Hamiltonian of the CFT, and thus the
perturbation can be absorbed by renormalizing the ``spin-wave velocity''.
Therefore, we do not have to consider such a perturbation explicitly.

On the other hand, the perturbation:
\begin{equation}
    g_{T^2} \int \mathrm{d}y \; \left[ T^2(w) + \bar{T}^2(\bar{w}) \right] \ ,
\end{equation}
which has the conformal dimension $(4,0)$ and $(0,4)$, is generally present in the effective theory of a lattice model\cite{reinickeAnalyticalNonanalyticalCorrections1987},
since it corresponds to the the breaking of the Lorentz invariance,
or equivalently continuous rotation symmetry
in the Euclidean space-time.

Although it is a renormalization group-irrelevant perturbation, it does
have nontrivial effects on energy levels:
\begin{align}
\label{T_mat0}
 \left( \delta \calE_n \right)^\text{(1)}_{T^2}&=
 2\pi g\frac{\pi^3}{L^3}\left[\left(\frac{c}{24}\right)^2+\frac{11c}{1440}+A(h,r)\right]\ ,
\end{align}
where $h$ is the conformal dimension 
of the primary field corresponding to the state $n$,
and $r$ is the level of the descendant, i.e.,
\begin{align}
    r=\sum_{i=1}^n k_i\  \text{if}\ \left|n\right\rangle=L^\text{(H)}_{-k_1} L^\text{(H)}_{-k_2} \cdots L^\text{(H)}_{-k_n}|h\rangle \quad\left(1 \leq k_1 \leq \cdots \leq k_n\right)\ ,
\end{align}
where $\{L^\text{(H)}_{n}\}$ are the Virasoro operators in the upper half plane as defined in \eqref{eq:vira}.
$A(h,r)$ are calculated by Reinicke~\cite{reinickeAnalyticalNonanalyticalCorrections1987}:
\begin{align}
A(0, r)&=\left(\frac{11}{30}+\frac{c}{12}\right) r\left(2 r^2-3\right), \quad r \neq 1\ ,\notag\\
A(h, r)&=(h+r)\left[\left(h-\frac{1}{6}-\frac{c}{12}\right)+\frac{r(2 h+r)(5 h+1)}{(h+1)(2 h+1)}\right], \quad h \neq 0\ .
\end{align}

For completeness, we also considered the $T\bar{T}$ bulk perturbation:
\begin{align}
    g_{T\bar{T}}\int \mathrm{d} y T\bar{T}(w,\bar{w})
\end{align}
Such perturbation has conformal dimension $(2,2)$, thus at the 1st order giving rise to FSC as
\begin{align}
    \delta \left( \calE_n \right)_{T\bar{T}}^\text{(1)}=\pi g_{T\bar{T}}\left(\frac{\pi}{L}\right)^3\left(\Delta+r-\frac{c}{24}\right)^2\ ,
\end{align}
which is at the same order as $T^2+\bar{T}^2$. For the XXZ chain, these two perturbation together give rise to the the leading-order or next-to-leading-order FSC\cite{PhysRevLett.58.771,LUKYANOV1998533,LUKYANOV2003323}. However, it does not present in the particular models we are considering in this paper (TF-Ising and three-state Potts)\cite{reinickeAnalyticalNonanalyticalCorrections1987}.
\subsection{Application to the critical Ising model}\label{sec:sum_ising}

We consider the standard quantum transverse-field Ising chain defined by
the Hamiltonian:
\begin{equation}
H = - \sum_{j=1}^{N-1} \sigma^z_j \sigma^z_{j+1} - \sum_{j=1}^N \sigma^x_j
    - \zeta_L \sigma^z_1 - \zeta_R \sigma^z_N\ .
\label{eq:IsingChain}
\end{equation}
This model can be solved exactly by the Jordan-Wigner transformation.
In the presence of the (bulk) longitudinal field, the model is no longer exactly
solvable.
Nevertheless, the above model with only the boundary longitudinal field can
be solved exactly\cite{campostriniQuantumIsingChains2015}.

\subsubsection{Free boundary (\texorpdfstring{$\zeta_R=\zeta_L=\zeta=0$}{lalphabeta})}

For $\zeta_R = \zeta_L =0$, we can identify the boundary condition as the free boundary
condition.
The exact ground-state energy for large $N$, expanded by powers of $1/N$, reads
\begin{align}\label{eq:Ising_FF}
E_0(\text{free},\text{free}) &= 2\left(\frac{1}{2}-\frac{1}{\pi}\right)-\frac{4}{\pi}N-\frac{\pi}{24 N}+\frac{\pi}{48 N^2}+\frac{-\frac{\pi}{96}-\frac{7 \pi^3}{23040}}{N^3}+\frac{\frac{\pi}{192}+\frac{7 \pi^3}{15360}}{N^4}+\cdots\ .
\end{align}
The $O(1/N^2)$ term, as discussed in \cite{2016_Piroli_SP_1}, can be eliminated by instead expanding \eqref{eq:Ising_FF} as
\begin{align}
    E_0(\text{free},\text{free}) &= 2\left(\frac{1}{2}\right)-\frac{4}{\pi}\left(N+1/2\right)-\frac{\pi}{24( N+1/2)}-\frac{7}{23040}\left(\frac{\pi}{ N+1/2}\right)^3+\cdots\ ,
    \label{eq:IsingFF}
\end{align}
where the "speed of light" in the TF-Ising model is set to $v=2$.

This suggests that the effective length $L$ of the system in the field theory should be identified with $(N+1/2)a$, so that
\begin{align}
    E_0(\text{free},\text{free}) &= 2\left(\frac{1}{2}\right)-\frac{4}{\pi}L-
    \frac{\pi v}{48 L}-\frac{7v }{2\times23040}\left(\frac{\pi}{L}\right)^3+\cdots\ .
\end{align}
In this paper, to compare the CFT result with the lattice model, we employed the convention that the lattice spacing $a$ is taken to $1$ for simplicity in the CFT expression\footnote{It should be noted that to recover the correct dimension in the continuum limit, one requires that $L=\left[N+(\delta L)_L+(\delta L)_R\right ]a$, where $a$ has the dimension of length. See Sec.~\ref{Sec_E_A} for an example.}. By Eq.~\eqref{eq:def_L}, the corrections to the length for the free boundary ($\zeta=0$) is
\begin{equation}
    \left(\delta L \right)_\text{free} = + \frac{1}{4}\ .
\end{equation}
By choosing the effective length in this way, the finite-size correction of $O(1/L^2)$,
which corresponds to the boundary operator $T$, is eliminated in Eq.~\eqref{eq:Ising_FF}. Furthermore, the energy eigenvalues of the first, second, and third excited states
are obtained as
\begin{align}
     E_1(\text{free},\text{free})
     &= E_0(\text{free},\text{free}) + \frac{\pi v}{2L} - \frac{v}{192}\left(\frac{\pi}{L}\right)^3
     + \cdots\ , \\
      E_2(\text{free},\text{free})
     &= E_0(\text{free},\text{free}) + \frac{3\pi v}{2L} - \frac{9 v}{64}\left(\frac{\pi}{L}\right)^3 + \cdots\ ,
\\
   E_3(\text{free},\text{free})
     &= E_0(\text{free},\text{free}) + \frac{2\pi v}{L} - \frac{7 v}{48}\left(\frac{\pi}{L}\right)^3 + \cdots\ ,   
\end{align}

The $O(1/L)$ terms in the energy eigenvalues are well known to match the
boundary CFT spectrum of the Ising model for the free boundary conditions.
$E_0$ has the universal correction to the ground-state energy
$- \pi v c/(24L)$ for the central charge $c=1/2$~\cite{cardyEffectBoundaryConditions1986,bloteConformalInvarianceCentral1986a}.
The $O(1/L)$ terms in the excitation energies are generally given as
$\pi v h/L$, where $h$ is the scaling dimension of the corresponding
boundary operator for the free boundary condition.
For the above lowest excited states, they are the
primary field with the scaling dimension $h=1/2$,
its first descendant with the scaling dimension $h=3/2$,
and the second descendant $L_{-2}$ of the identity operator with the scaling dimension $h=2$.

We can furthermore identify the $O(1/L^3)$ terms as the energy shift due to the
bulk perturbation $T^2 + \bar{T}^2$.
Evaluating Eq.~\eqref{T_mat0} for the Ising model, we find the CFT prediction:
\begin{align}
 \left( \delta \calE_0 \right)^{\text{(1)}}_{T^2} &=
\langle 0,0 |H_{T^2}|0,0\rangle =2g_{T^2} v\frac{\pi^4}{L^3}\frac{c(22+5c)}{2880}\ ,
\label{eq:E0_T2}
\\
 \left( \delta \calE_1 \right)^{(1)}_{T^2} -
  \left( \delta \calE_0 \right)^{(1)}_{T^2} 
 &=
\langle 1/2,0 |H_{T^2}|1/2,0 \rangle -\langle 0,0 |H_{T^2}|0,0\rangle =
2g_{T^2} v\frac{\pi^4}{L^3}\frac{7}{48}\ ,\\
 \left( \delta \calE_2 \right)^{(1)}_{T^2} -
  \left( \delta \calE_0 \right)^{(1)}_{T^2} 
 &=
\left<1/2,1 |H_{T^2}|1/2,1\right>-\left< 0,0 |H_{T^2}|0,0\right>= 2g_{T^2}v\frac{\pi^4}{L^3}\frac{63}{16}\ ,\\
 \left( \delta \calE_3 \right)^{(1)}_{T^2} -
  \left( \delta \calE_0 \right)^{(1)}_{T^2} 
 &=
\left< 0,2 |H_{T^2}|0,2\right>-\left< 0,0 |H_{T^2}|0,0\right> =2g_{T^2}v\frac{\pi^4}{L^3}\frac{49}{12}\ ,
\label{eq:E3_T2}
\end{align}

Since the bulk perturbation should be independent of the boundary condition,
we expect that the coupling constant is identical between OBC and PBC:
\begin{align}
\label{g_same}
    g_i^\text{OBC}=g_i^\text{PBC}\ ,
\end{align}
as long as the same normalization is taken in both boundary conditions.
From the finite-size spectrum of the critical Ising model~\eqref{eq:IsingChain}
with the PBC, we can extract the coupling constant~\cite{reinickeAnalyticalNonanalyticalCorrections1987}
\begin{equation}\label{g_T}
    g_{T^2} = - \frac{1}{56 \pi}\ .
\end{equation}
Using the same coupling constant for Eqs.~\eqref{eq:E0_T2}--\eqref{eq:E3_T2},
we can exactly reproduce the results of the lattice model.
This confirms that the $O(1/L^3)$ corrections are indeed due to the bulk $T^2 + \bar{T}^2$ perturbation and the identity~\eqref{g_same}.
Eq.~\eqref{g_same} will be verified again in the three-state Potts model,
where we have the bulk $X\bar{X}$ perturbation.

\subsubsection{Fixed boundary conditions}
A natural implementation of the fixed boundary condition for the Ising model~\eqref{eq:IsingChain} is to simply restrict the spin at the end of
the chain to $\uparrow$ ($\sigma^z=+1$) or $\downarrow$ ($\sigma^z=-1$). The energy spectrum of the Ising model with such fixed boundary condition is particularly simple, thanks to the Kramers-Wannier duality (Appendix \ref{app:K-W}):
\begin{align}\label{eq:KW_parallel}
E_0(N+1, \uparrow, \uparrow) = E_0(N+1,\downarrow,\downarrow)
&= E_0(N,  \text{free},  \text{free} )\ ,
\\
E_0(N+1, \uparrow, \downarrow) = E_0(N+1,\downarrow,\uparrow)
&= E_1(N, \text{free},  \text{free} )\ .\label{eq:KW_antiparallel}
\end{align}
where the first parameter in the parentheses indicates total number of sites, including those fixed ones.
This suggests that in this implementation of the fixed boundary conditions, the correction to the length for the fixed boundary conditions is
\begin{align}
    \left(\delta L \right)_{\uparrow/\downarrow} = - \frac{1}{4} \ .
\end{align}
It should be noted however that the correction to the effective length is non-universal,
and is a function of the boundary field $\zeta$~\cite{campostriniQuantumIsingChains2015}.
As is evident from the identities, the energy eigenvalues under the fixed
boundary conditions also has $O(1/L^3)$ corrections due to the bulk $T^2 + \bar{T}^2$
perturbation.

Only restricting the spin at one end of the chain while leaving the other end free corresponds to the Brascamp-Kunz boundary condition \cite{izmailianIsingModelMixed2009,brascampZeroesPartitionFunction1974}. In that case, the ground-state energy reads
\begin{align}
    E_0(\uparrow/\downarrow,\text{free}) &= 2\left(\frac{1}{2}\right)-\frac{4}{\pi}N+\frac{\pi}{12 N}+\frac{1}{2880}\left(\frac{\pi}{ N}\right)^3+\cdots\ .
\end{align}
We notice that there is no $O(1/N^2)$ term, which can be understood as
\begin{align}
    L=N+(\delta L)_{\uparrow/\downarrow}+(\delta L)_\text{free}=N\ .
\end{align}
And the $O(1/L^3)$ correction is again due to the $T^2+\bar{T}^2$ perturbation:
\begin{align}
    \left( \delta \calE_0 \right)^{(1)}_{T^2} 
 &=
\langle \frac{1}{16},0 |H_{T^2}| \frac{1}{16} ,0 \rangle =
2g_{T^2} v\frac{\pi^4}{L^3}\left(-\frac{7}{1440}\right)\ ,
\end{align}
where $g_{T^2}$ takes the same value as in Eq.~\eqref{g_T}. Therefore, for all the conformally invariant boundary conditions, we may apply the same effective Hamiltonian:
\begin{align}
    H_\text{eff}=H^\text{CFT}-\frac{1}{56\pi} \int \mathrm{d} x \; \left[T^2(x)+\bar{T}^2(x)\right]
\end{align}
to obtain the leading order (LO) finite-size correction.

\subsubsection{Boundary field perturbation to the free boundary}
Following Ref.~\cite{campostriniQuantumIsingChains2015},
we consider the small boundary field $\zeta_L = \pm \zeta_R = \zeta$
in the scaling limit $L \to \infty$ keeping
\begin{align}
    \zeta_b = \zeta L^{1/2}
\end{align}
constant.
In this limit, the finite-size corrections to the energy spectrum
are due to the boundary field perturbations.
Since the free boundary condition is invariant under the spin-flip parity,
the first-order correction in the boundary field vanishes.
Thus, the leading finite-size correction is in the second order of the boundary
fields.
Since the boundary scaling dimension of the boundary spin operator is $1/2$,
we use Eq.~\eqref{eq:boundary2nd_h0.5} to obtain
\begin{align}
\left( \delta \calE_0 \right)^{(2)}_\zeta &=
  - 2 v {c_\sigma}^2 \zeta^2
  \left\{
    \text{const.} + \log{L} \pm  \frac{\pi}{2} 
  \right\}\ ,
\label{eq:IsingBF_CFT}
\end{align}
where $-$ of the double sign
corresponds to the parallel boundary field $\zeta_L = \zeta_R =\zeta$
and $+$ to the antiparallel boundary field $\zeta_L = - \zeta_R = \zeta$.
$c_\sigma$ is the non-universal renormalization factor of the spin operator at the boundary
site, which is defined by
\begin{align}
    \langle \sigma^z_1(\tau) \sigma^z_1(0) \rangle \sim \frac{{c_\sigma}^2}{v \tau}
\end{align}
in the limit of large $\tau$.

On the other hand, in the scaling limit,
from the exact solution of the Ising model with the boundary fields 
we find (see Appendix~\ref{app: magnetic} for the derivation)
\begin{align}\label{eq:E0_mag_lat}
E_0(\zeta_L=\zeta,\zeta_R=\pm\zeta)=E_0(\text{free},\text{free}) \mp \frac{v}{2} \zeta^2-\frac{v}{\pi}\left(\log 2-\log \frac{v \lambda}{ 2 L}\right) \zeta^2\ ,
\end{align}
where $\lambda \to +0$ is a UV cutoff.
Indeed, this agrees with the perturbed CFT result~\eqref{eq:IsingBF_CFT},
except for the ambiguity in the constant.
By the comparison of two formulae, we can also determine the renormalization factor:
\begin{align}
    {c_\sigma}^2 &= \frac{1}{2\pi}\ ,
\end{align}
for the spin operator at the boundary site.
\subsection{Application to the three-state Potts model}

Now let us apply our results to the quantum three-state Potts chain~\cite{saleurRelationsLocalHeight1989,cardyBoundaryConditionsFusion1989,affleckBoundaryCriticalPhenomena1998}:
\begin{align}
    H=-\sum_i\left(  M_i+M_i^{\dagger}+R_i^{\dagger} R_{i+1}+R_i R_{i+1}^{\dagger} \right)
    - \left( \zeta_L R_1 + \zeta_L^* R_1^\dagger \right) - \left( \zeta_R R_N + \zeta_R^* R_N^\dagger \right)\ ,
    \label{eq:PottsH}
\end{align}
where 
\begin{align}
    M=\left(\begin{array}{ccc}
0 & 1 & 0 \\
0 & 0 & 1 \\
1 & 0 & 0
\end{array}\right) , \quad R=\left(\begin{array}{ccc}
e^{2 \pi i / 3} & 0 & 0 \\
0 & e^{4 \pi i / 3} & 0 \\
0 & 0 & 1
\end{array}\right)\ .
\end{align}
The boundary fields $\zeta_{L,R}$ are generally complex.

The numerical results on the three-state Potts model in the following were obtained by
DMRG of finite-size chains.
We have used the DMRG function in ITensors library, which calculates the ground state in matrix-product-state form, with bond dimensions up to 400
to study finite-size systems up to $300$ sites. The first excited state is obtained as the lowest energy state in the subspace orthogonal to the ground state.
The numerical error in the energy is estimated to be less than $1.4\times 10^{-4}$.
While there are limitations of DMRG due to finite bond dimensions,
we are dealing with sufficiently small systems where the DMRG
results are essentially exact.

\subsubsection{Free boundary (\texorpdfstring{$\zeta_R=\zeta_L=\zeta=0$}{lalphabeta})}

For $\zeta_R = \zeta_L =0$, we can identify the boundary condition as the free boundary
condition.
Using the exact solution for the $O(1)$ boundary energy~\cite{10.1143/PTP.102.39}, 
the ground-state energy for large $N$ can be written as
\begin{align}
E_0(\text{free},\text{free}) =
 \left(-\frac{4}{3}-\frac{2\sqrt{3}}{\sqrt{\pi}} \right) N
 + \frac{3\sqrt{3}}{4}-1
 + \frac{\pi v}{24 N} c + \cdots\ ,
\end{align}
where $c=4/5$ is the central charge of the three-state Potts model and $v=\frac{3\sqrt{3}}{2}$.
As in the case of the Ising model,
we should be able to eliminate the finite-size correction $\propto 1/N^2$
by using the
appropriate effective length $L = N a + 2 \left( \delta L \right)_\text{free}$.

The three-state Potts model has 5 bulk operators which are RG-relevant.
Among them, the spin operators $\sigma_{1,2} \bar{\sigma}_{1,2}$ with the scaling dimension
$\Delta_\sigma = 2/15$ and $\psi_{1,2} \bar{\psi}_{1,2}$
with the scaling dimension $\Delta_\psi=4/3$
are charged under the $\mathbb{Z}_3$ symmetry, and thus are forbidden if
the $\mathbb{Z}_3$ symmetry is imposed on the bulk.
The other relevant operator $\epsilon \bar{\epsilon}$
with the scaling dimension $\Delta_\epsilon=4/5$
corresponds to the thermal perturbation; it is absent in the effective theory of the
critical Potts model, thanks to the fine-tuning to the critical point.
Thus, the leading finite-size correction due to bulk operators
is induced by the operator $X\bar{X}$ with the scaling dimension $14/5$.

The effective field theory for the free boundary condition on both ends
is given by
\begin{align}\label{eq:H_eff_Potts_free}
 H_\text{eff}=H_\text{free-free}+\gamma\int_{0}^{L} \mathrm{d}v\, X\bar{X}(0, v) +\beta_2 \int_{0}^{L} \mathrm{d}v\,\left(T^2(0, v)+\bar{T}^2(0, v)\right)+\cdots\ ,
\end{align}
where we adopt the standard normalization~\eqref{eq:bulkOP_normalization}
of the bulk operators.

We note that the leading boundary perturbations on the free boundary
condition are $\psi$ and $\bar{\psi}$ with the scaling dimension $2/3$.
However, they correspond to the boundary field breaking the $\mathbb{Z}_3$ symmetry,
and are absent in the lattice model with free boundaries preserving the $\mathbb{Z}_3$
symmetry.
The next leading boundary perturbation is the
boundary $T$ perturbation with the scaling dimension $2$, which
can be absorbed by renormalizing the system size as discussed in Sec.~\ref{sec_L}.
The next leading boundary perturbation is the level-$3$ descendant of
the identity operator with the scaling dimension $3$,
which would give a $O(L^{-3})$ correction which is smaller than
any of the corrections due to bulk perturbations discussed above.

\begin{figure}[tb]
  \centering
\includegraphics[keepaspectratio, width=86mm]{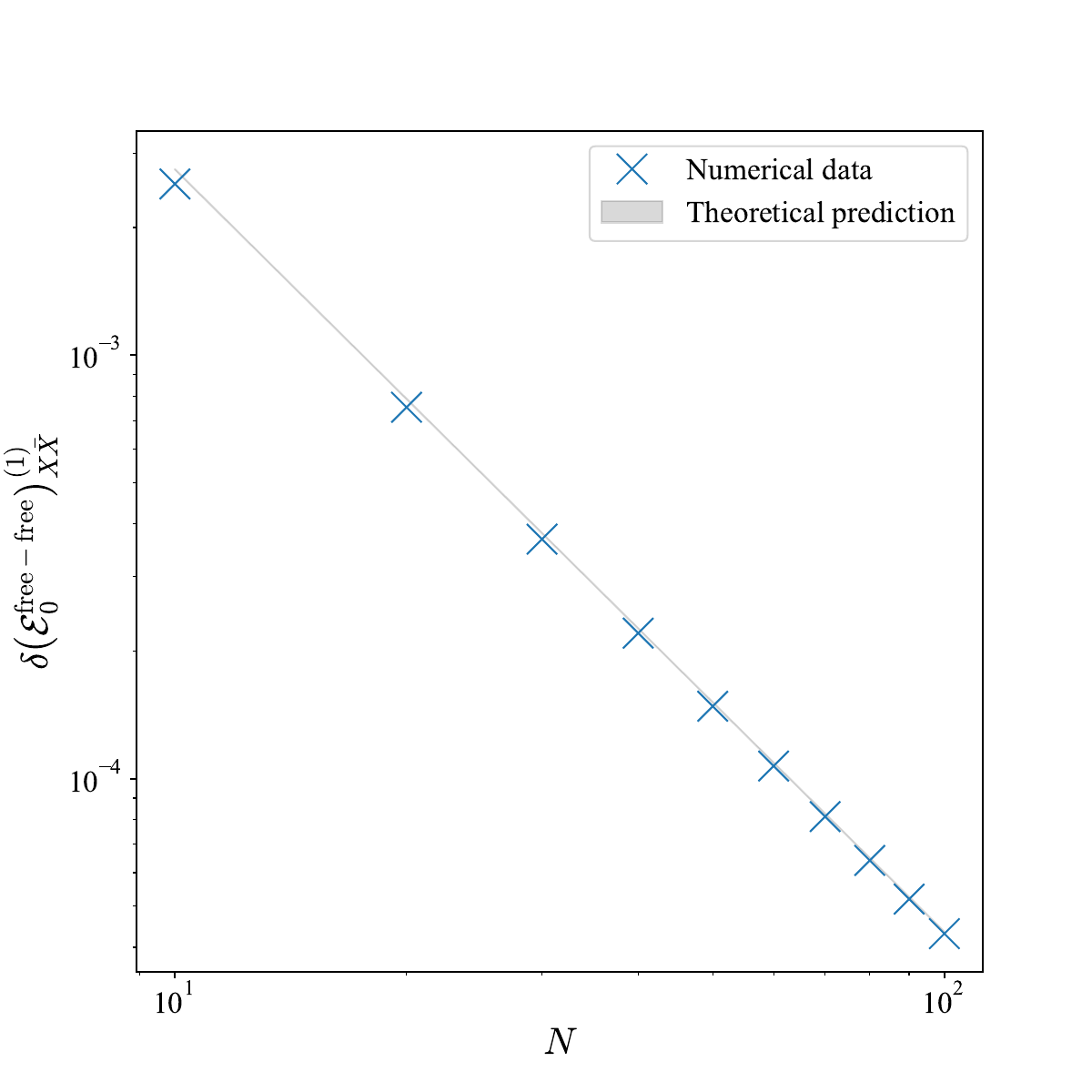}
  \caption{Comparison between the numerical extracted $\delta \left( \calE_0^{\text{free-free}} \right)^{(1)}_{X\bar{X}}$ for the three-state Potts model and the theoretical prediction in \eqref{eq:E0_bulk_free}. Error from \eqref{eq:gamma} is indicated by thickness.}
  \label{fig:3_free}
\end{figure}

Using Eq.~\eqref{eq:E0_bulk_1st}, the leading order FSC to the ground state
is thus given by
\begin{align}\label{eq:E0_bulk_free}
    \delta \left( \calE_0^{\text{free-free}} \right)_{X\bar{X}}^\text{(1)} = \gamma v\left(\frac{\pi}{L}\right)^{9/5}{}^{\text{free}}B_{X\bar{X}}^{\mathds{1}}\frac{\sqrt{\pi}\Gamma(-9/10)}{2^{14/5}\Gamma(-2/5)}\ .
\end{align}
The bulk-boundary OPE coefficient is known as~\cite{runkelStructureConstantsDseries2000}\footnote{Since we are using a different normalization from~\cite{runkelStructureConstantsDseries2000}, our OPEs are also different. Such difference will be shown explicitly as in \eqref{eq:XX_1st_free}.}
\begin{align}\label{eq:XX_1st_free}
    {}^{\text{free}}B_{X\bar{X}}^{\mathds{1}} &= \frac{^{(\omega)} B_{(3,1)\overline{(3,1)}}^\mathds{1}}{\sqrt{C_{(3,1)\overline{(3,1)}\ (3,1)\overline{(3,1)}}^\mathds{1}}}=\sqrt{\frac{S_{\varepsilon}}{S_\mathds{1}}}=\sqrt{\frac{\sin(2\pi/5)}{\sin(\pi/5)}}.
\end{align}
The coupling constant $\gamma$ for the bulk perturbation $X\bar{X}$ appearing
in Eq.~\eqref{eq:H_eff_Potts_free}
was extracted from the PBC spectrum in Ref.~\cite{reinickeAnalyticalNonanalyticalCorrections1987}
as
\begin{align}\label{eq:gamma}
    \gamma &= 0.009 237 (7).
\end{align}
Using these values, Eq.~\eqref{eq:E0_bulk_free} indeed agrees very well
with the numerical results as shown in Fig.~\ref{fig:3_free}.

The next order correction after the above $O(L^{-9/5})$ due to the bulk $X\bar{X}$
is expected to be $O(L^{-2})$ due to the boundary $T$ operator, which can be absorbed
by renormalizing the effective system size as discussed in Sec.~\ref{sec:T_per}.
After subtracting the predicted $O(L^{-9/5})$ correction, however,
the numerically obtained ground-state energy does not show the
generally expected $L^{-2}$ correction.
This suggests that the length correction vanishes
\begin{equation}
    \left(\delta L\right)_\text{free}=0
    \label{eq:deltaL_free}
\end{equation}
for the free boundary condition for the model~\eqref{eq:PottsH}.
(Numerically, this holds up to $|\left(\delta L\right)_\text{free}|$ $<0.02$ 
and we expect that Eq.~\eqref{eq:deltaL_free} is exact.)
In the absence of the $L^{-2}$ correction, the next leading correction
seems to be $O(L^{-13/5})$, which is attributed
to the second order perturbation from the bulk $X\bar{X}$.

\subsubsection{Fixed/Mixed boundary conditions}\label{Sec_3-state_fixed}

Similarly to the Ising model, the boundary magnetic field is a relevant perturbation
to the free boundary condition.
However, in the three-state Potts model, the boundary magnetic field will drive the
boundary condition to qualitatively different ones, depending on its complex phase.
For real $\zeta_{L,R}>0$, the boundary magnetic favors the single state $R=1$.
Hence the induced boundary condition is ``fixed'' to A ($R=1$).
Similarly, $\zeta_{LR} \propto e^{\pm 2\pi i/3}$ drives the boundary condition
to fixed to B or C ($R= e^{\pm 2 \pi i/3}$).
On the other hand, if $\zeta_{L,R}$ is real and negative, it favors
two states B and C equally, resulting in the ``mixed'' boundary condition
of B and C. 

As in the Ising case, we can naturally realize fixed/mixed boundary conditions by restricting the Hilbert space of boundary spin to a subspace with a proper subset of possible spin directions. Once again, thanks to the explicit duality in the three-state Potts model \cite{affleckBoundaryCriticalPhenomena1998}, the energy spectrum with fixed boundary conditions read
\begin{align}\label{eq:3-state_same}
E_0(N+1, \text{A}, \text{A})= E_0(N+1,\text{B},\text{B}) = E_0(N+1,\text{C},\text{C})
&= E_0(N, \text{free}, \text{free} )\ ,
\\\label{eq:3-state_mixed}
E_0(N+1, \text{A}, \text{B}) = E_0(N+1,\text{A},\text{C}) = E_0(N+1,\text{B},\text{C}) = \ldots 
&= E_1(N, \text{free}, \text{free} )\ .
\end{align} 
In fact, the LO FSC to the ground state is indeed the same as Eq.~\eqref{eq:E0_bulk_free}:
\begin{align}
    \delta \left( \calE_0^{\text{A-A}} \right)^{(1)}_{X\bar{X}} = \gamma v\left(\frac{\pi}{L}\right)^{9/5}{}^{\text{A}}B_{X\bar{X}}^{\mathds{1}}
    \frac{\sqrt{\pi}\Gamma(-9/10)}{2^{14/5}\Gamma(-2/5)} ,
\end{align}
since
\begin{align}
    {}^{\text{A}}B_{X\bar{X}}^{\mathds{1}} &= \frac{^{(1,2)} B_{(3,1)\overline{(3,1)}}^\mathds{1}}{\sqrt{C_{(3,1)\overline{(3,1)}\ (3,1)\overline{(3,1)}}^\mathds{1}}}=\sqrt{\frac{S_{\varepsilon}}{S_\mathds{1}}}=\sqrt{\frac{\sin(2\pi/5)}{\sin(\pi/5)}}={}^{\text{free}}B_{X\bar{X}}^{\mathds{1}} .
\end{align}

\begin{figure}[tb]
  \centering
\includegraphics[keepaspectratio, width=86mm]{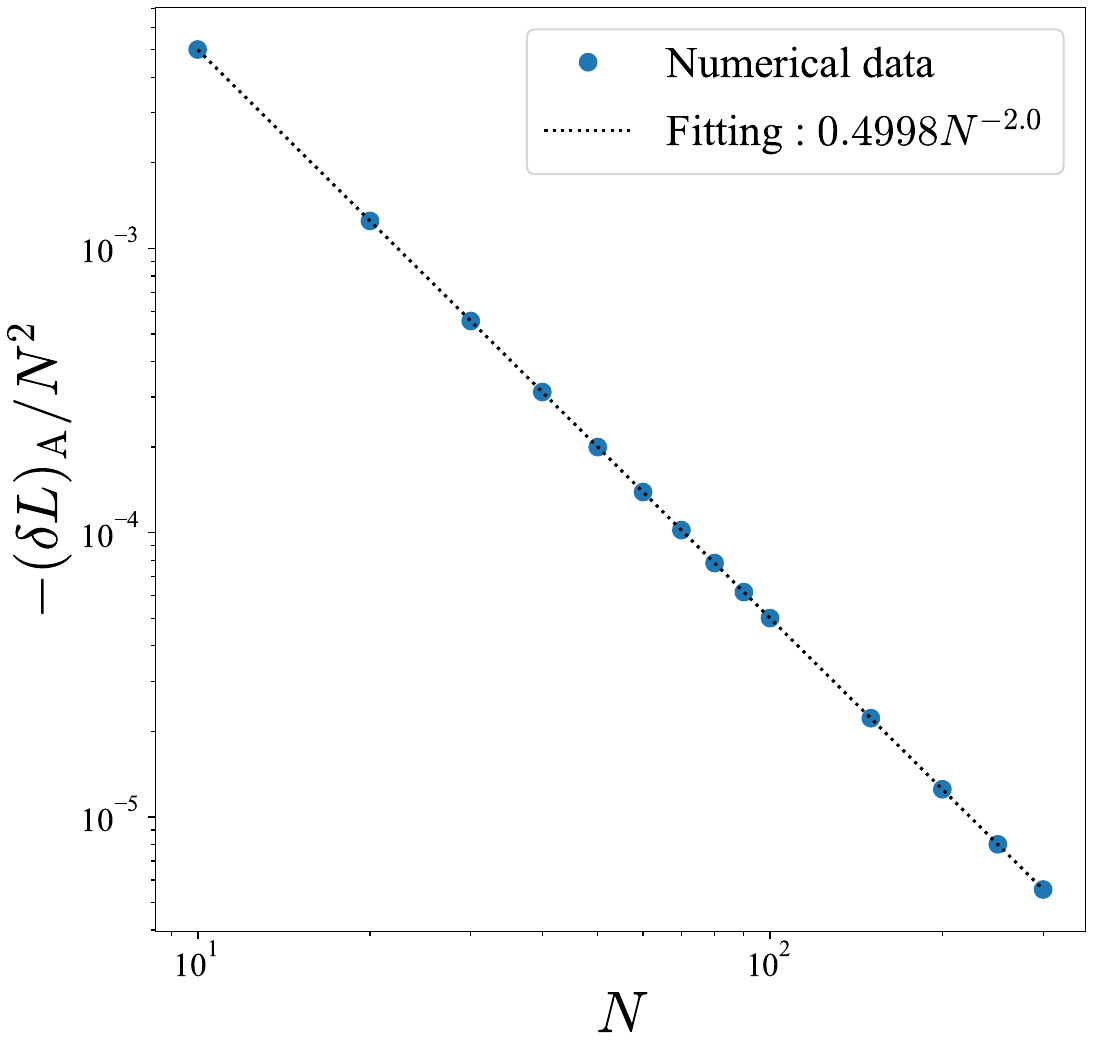}
  \caption{$-(\delta L)_\text{A}/N^2$ obtained from the first-order correction to the ground-state energy for A-A boundary condition for the three-state Potts model. The numerical fitting suggests $(\delta L)_\text{A}=-0.5$ up to three digits. }
  \label{fig:L_A}
\end{figure}

The duality relation~\eqref{eq:3-state_same} also implies
\begin{equation}
    \left( \delta L \right)_\text{A,B,C} = \left(\delta L \right)_\text{free} - \frac{1}{2}.
\end{equation}
This together with the numerical finding~\eqref{eq:deltaL_free} implies
\begin{align}
    \left( \delta L \right)_\text{A,B,C}=-\frac{1}{2},
\end{align}
which is also verified by the direct numerical calculation
as shown in Fig.~\ref{fig:L_A}. 

To realize the mixed boundary conditions, we restrict the Hilbert space of the boundary spin to a limited one. This is known as the "blob" boundary conditions \cite{jacobsenConformalBoundaryLoop2008}. For example, for AB or BC boundary conditions, the local operators on the edges are replaced with projected ones, such as
\begin{gather*}
    M_\mathrm{AB} = \begin{pmatrix}
 0& 0 & 0 \\
 0&  0& 1 \\
 0& 0 &  0\\
\end{pmatrix},\quad 
    M_\mathrm{BC} = \begin{pmatrix}
 0& 1 & 0 \\
 0&  0& 0 \\
 0& 0 &  0\\
\end{pmatrix}. \\
    R_\mathrm{AB} = \begin{pmatrix}
 0& 0 & 0 \\
 0&  e^{4\pi i/3}& 0 \\
 0& 0 &  1\\
\end{pmatrix},\quad
    R_\mathrm{BC} = \begin{pmatrix}
 e^{2\pi i/3}& 0 & 0 \\
 0&  e^{4\pi i/3}& 0 \\
 0& 0 &  0\\
\end{pmatrix}.
\end{gather*}

\noindent For these boundary conditions, in addition to \eqref{eq:H_eff_Potts_free}, the effective Hamiltonian also contains irrelevant boundary perturbation from the $X$ operator living on the mixed boundary condition:
\begin{align}\label{Heff3state}
H_\text{eff}=H_{\text{AB}-\calB}+g^L_X X^{\text{(AB)(AB)}}(0,0)
+\gamma \int_{0}^{L}  X\bar{X}(0,v)\,\mathrm{d}{v}+ 
\cdots\ ,
\end{align}
where $\calB$ is the boundary condition on the right end. (If $\calB$ is also a mixed boundary condition, there is also a correction proportional to the perturbation $g^R_X$ on the right.) 
Since the boundary $X$ operator has scaling dimension $h_X=7/5$, the first order perturbation, according to \eqref{eq:bound_1st}, due to the boundary $X$ perturbation to
the mixed boundary condition on the left reads
\begin{align}
\label{eq:X_1st}
    \delta \left(\calE_n^{\text{AB}-\calB}\right)^{(1)}_{X} =
    g^L_X v\left(\frac{\pi }{L}\right)^{7/5} &
     C^{(\text{AB})(\text{AB})\calB}_{nXn} ,
\end{align}
where $\calB$ is the boundary condition on the right end.
(If $\calB$ is also a mixed boundary condition, there is also a correction proportional to the perturbation $g^R_X$ on the right.) 

\begin{figure}[tb]
  \centering
\includegraphics[keepaspectratio, width=86mm]{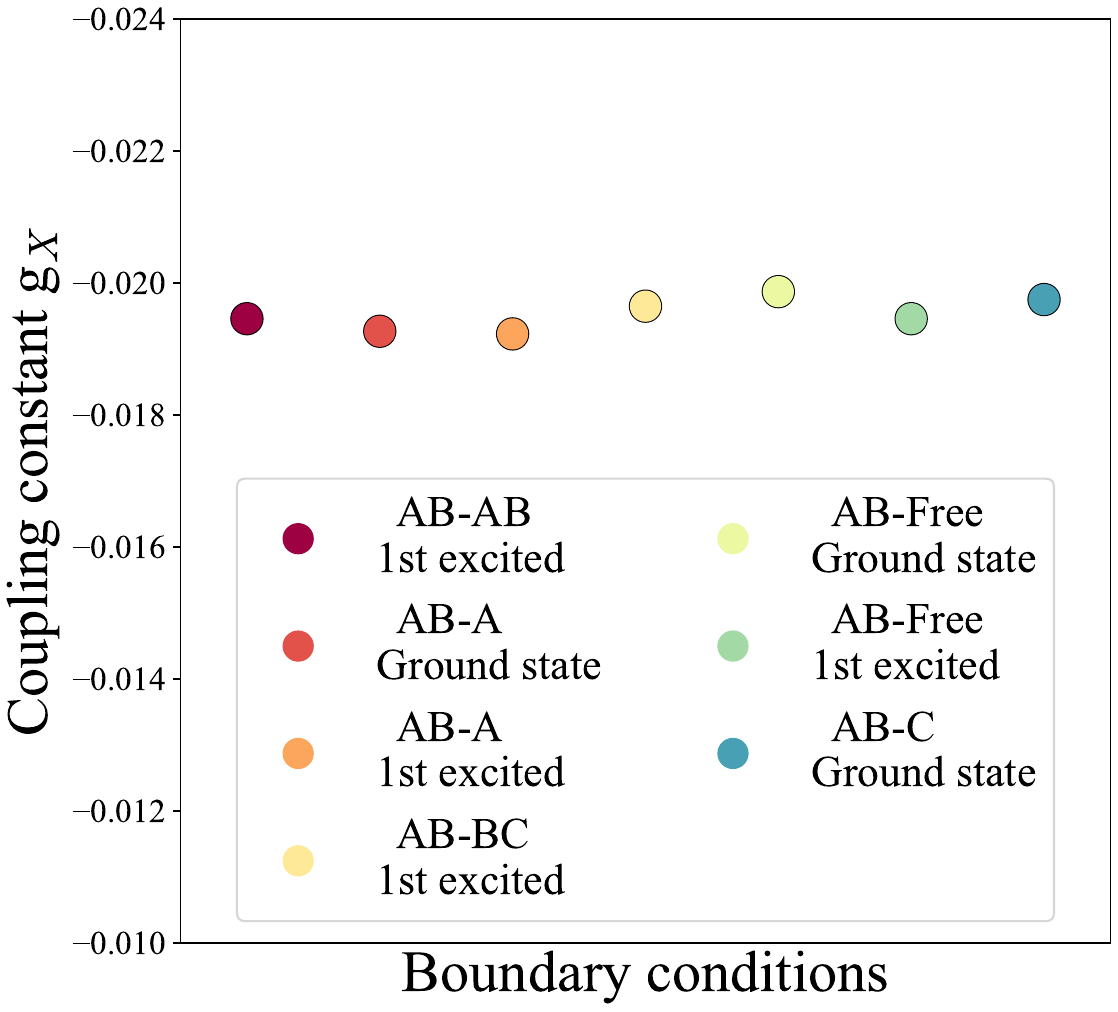}
  \caption{Estimated coupling constant $g_X$ from the numerical data for several combinations
  of boundary states involving the mixed boundary condition.
  Expected values are labeled by their level of excitation and the combination of
  boundary conditions. Data in this figure suggest that $g_X=0.0195(3)$. Cases where a first-order perturbation is prohibited by the fusion rule are not shown.}
  \label{gX}
\end{figure}

For different pairs of boundary conditions involving the mixed boundary conditions,
we find different values of the $O(L^{-7/5})$ corrections at various energy levels.
In Fig.~\ref{gX}, we show the estimated values of the coupling constant $g_X$ of
the boundary perturbation $X$ on the mixed boundary conditions,
obtained by fitting several $O(L^{-7/5})$ FSC at different energy levels with also different combinations of the boundary conditions.
Since $g_X$ is intrinsic to the mixed boundary condition on the lattice model,
its value must be unique for different combinations.
In fact, all the different estimates are consistent with the unique value:
\begin{align}
  g_X^L=g_X^R=g_X=0.0195(3)\ .  
\end{align}
This serves as an evidence supporting the validity of our analysis.

Nevertheless, due to the fusion rule, Eq.~\eqref{eq:X_1st} vanishes for certain combinations of boundary conditions at some energy levels. For instance, let us consider the leading order FSC to the ground-state energy when the same mixed boundary condition is applied on each end, i.e., $\delta\left(\calE_0^\text{AB-AB}\right)$. With this diagonal boundary condition, the ground state corresponds to the vacuum, and thereby the LHS of Eq.~\eqref{eq:X_1st} is prohibited by the fusion rule as $C^{(\text{AB})(\text{AB})\calB}_{\mathds{1}X\mathds{1}}=0$.

This leaves the leading order corrections to the ground-state energy to the next possible order, $O(1/L^{-9/5})$. This time, at this order, one encounters a combination of the 1st order one due to the bulk $X\bar{X}$ perturbation and
the second order one due to the boundary $X$, both scale as $L^{-9/5}$. Following Eqs.~\eqref{eq:E0_bulk_1st} and~\eqref{eq:bo_2nd}, the leading order FSC reads
\begin{align}\label{eq:1.8_2nd}
    \delta\left(\calE_0^\text{AB-AB}\right)=\delta \left(\calE^{\text{AB-AB}}_0\right)^\text{(1)}_{X\bar{X}} + \delta \left(\calE^{\text{AB-AB}}_0\right)_X^\text{(2)}&= \left(\frac{\pi}{L}\right)^{9/5}
    \left[{}^{\text{AB}}B_{X\bar{X}}^{\mathds{1}}\frac{\sqrt{\pi}\Gamma(-9/10)}{2^{14/5}\Gamma(-2/5)} \gamma
    +\frac{21}{20}g_X^2\right]\ ,
\end{align}
where
\begin{align}
    {}^{\text{AB}}B_{X\bar{X}}^{\mathds{1}} &= \frac{^{(3,3)/(3,4)} B_{(3,1)\overline{(3,1)}}^\mathds{1}}{\sqrt{C_{(3,1)\overline{(3,1)}\ (3,1)\overline{(3,1)}}^\mathds{1}}}=-\frac{3-\sqrt{5}}{2}\sqrt{\frac{\sin(2\pi/5)}{\sin(\pi/5)}}\ .
\end{align}

\begin{figure}[tb]
  \centering
\includegraphics[keepaspectratio, width=86mm]{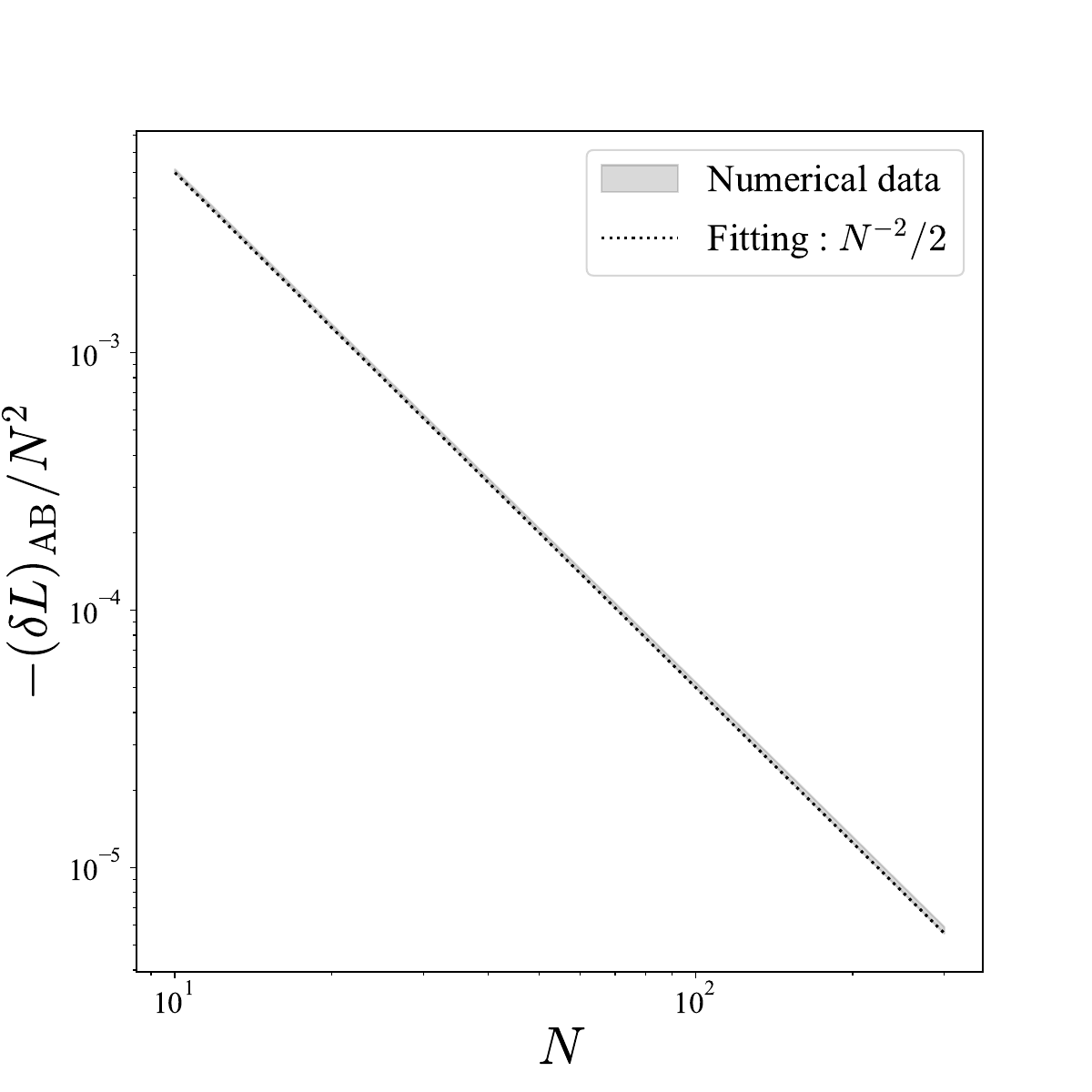}
  \caption{$-(\delta L)_\text{AB}/N^2$ obtained from the next leading-order correction to the ground-state energy for AB-AB boundary condition. Error from the uncertainty in $g_X$ is indicated by the thickness. The numerical fitting suggests $(\delta L)_\text{AB}=-0.5$ as well up to 3 digits. }
  \label{fig:L_AB}
\end{figure}

The next-to-leading order (NLO) correction to the ground-state energy is $O(L^{-2})$ due to the boundary
$T$ perturbation, which can be absorbed by renormalization of the system size
as discussed in Sec.~\ref{sec_L}.
The numerical result (Fig.~\ref{fig:L_AB}) suggests
\begin{align}
    (\delta L)_\text{AB,BC,CA}=(\delta L)_\text{A,B,C}=-\frac{1}{2}\ .
\end{align}


\section{Boundary Conformal Perturbation Theory and finite-size correction}
\label{sec_perturbation}

In this Section, we develop a general theory of FSC of energy levels of a CFT with OBC,
in the presence of both bulk and boundary perturbations.
Although we will mainly focus on the irrelevant perturbation in this paper, the results in this section can also be used to discuss relevant perturbation at the perturbative region.

\subsection{Finite-size spectrum under PBC}

Let us first briefly review the energy spectrum of a finite-length one-dimensional quantum critical system described by a CFT under PBC~\cite{cardy1984conformal,cardy1986operator}.
If the effective field theory were exactly given by a CFT, the finite-size effective quantum Hamiltonian and its spectrum read
\begin{align}
    H_\text{PBC}&=\frac{2\pi v}{N}(L_0+\bar{L}_0-\frac{c}{12})\ ,\notag\\
    E_n-\varepsilon_0N& = \frac{2\pi v}{N}(x_n-\frac{c}{12})\ ,
\end{align}
where $N$ is the number of lattice sites, $v$ is a characteristic velocity, $L_n,\bar{L}_n$ are the Virasoro modes, $\varepsilon_0$ is the bulk energy per site,
$x_n$ is the scaling dimension for $n$-th excited state and $c$ is the central charge.
This indeed gives the exact asymptotic spectrum in the limit $N \to \infty$.

However, for a more precise description of the system with a finite $N$, we need to include the perturbations to the CFT,
which corresponds to the renormalization group (RG) fixed point.
The quantum Hamiltonian of the effective field theory, including the perturbations, read
\begin{align}\label{eq_HPBC}
    H=H_\text{PBC}+\sum_i g_i\int\mathrm{d}x\Phi_i(x)\ ,
\end{align}
where the integration is performed over the spatial direction, and the sum is performed on all the local operators allowed by the symmetry of the Hamiltonian in the CFT.
The coupling constant $g_i$ grows under RG if the corresponding operator $\Phi_i$ is RG-relevant, namely if its scaling dimension
is smaller than the space-time dimension $2$.
At the critical point of the lattice model, all the relevant coupling constants $g_i$ should be tuned to zero.
However, the energy spectrum is still affected by the remaining irrelevant perturbations as
\begin{align}
    E_n(N)=\varepsilon_0N+\frac{2\pi v}{N}(x_n-\frac{c}{12})+\sum_i a_n^i                                    N^{-b^i_n}\ .
\end{align}
The third term is called the finite-size correction (FSC). The coefficients ${a^i_n}$ and $b^i_n$ can be explained by the effective Hamiltonian~\eqref{eq_HPBC}.
A detailed analysis of such irrelevant perturbations can be found in \cite{reinickeAnalyticalNonanalyticalCorrections1987}. We also refer the reader to \cite{henkelConformalInvarianceCritical1999} for a review of this topic.

\subsection{Boundary conformal field theory: a review }

Let us first review the spectrum of a CFT on an open strip of width $L$.
Each side of the strip is characterized by a conformally invariant boundary condition.
This corresponds to the spectrum of a lattice model of $N$ sites with open boundaries in the limit $N \to \infty$.
While the length $L$ in field theory is determined by $L \approx N$, there is a subtlety in the exact relation which we will discuss later.
We denote the temporal and spatial coordinates by $x\in(-\infty,\infty)$ and $y\in[0,L]$. The boundary conditions $\alpha$ and $\beta$ are now located at $y=0$ and $y=L$. In this geometry, like the usual cylinder geometry, we have the Hamiltonian for OBC:
\begin{align}
    H_{\alpha\beta}=\int_0^L(T(w)+\bar{T}(\bar{w}))\frac{\mathrm{d}y}{2\pi}\ ,
\end{align}
where $w=x+iy$\footnote{For simplicity of the expressions, when we are not referring to any particular lattice model, we set $v=1$ in the field theory result.}.
We can map this infinite long strip onto the upper half of the complex plane through a conformal mapping
\begin{align}
    z=\exp\left(\frac{\pi w}{L}\right)\ .
\end{align}
In the following, we will mainly work with this geometry. The boundary conditions are now located at different sides of the real axis. More specifically, we have $\alpha$ boundary condition on $\text{Re}(z)>0$ and $\beta$ boundary condition on $\text{Re}(z)<0$.

The conformal boundary condition implies that $T(z)=\bar{T}(\bar{z})$ for $\text{Im}(z)=0$. This condition allows us to do the analytical continuation for $T(z)$ to the lower half plane
\begin{align}
    \mathbb{T}(z)\coloneqq\sum_{n \in \mathbb{Z}} L_n^\text{(H)} z^{-n-2}=\left\{\begin{array}{lll}
T(z) & \text { for } & \text{Im} (z) \geq 0 \\
\bar{T}(\bar{z}) & \text { for } & \text{Im} (z)<0
\end{array}\right.\ ,
\end{align}
where $L_n^{(H)}$ is the Virasoro operator in the upper half plane:
\begin{align}\label{eq:vira}
L_n^\text{(H)}&\coloneqq\frac{1}{2 \pi i} \int_{C_+} z^{n+1} T(z) \,\mathrm{d}{z}-\frac{1}{2 \pi i} \int_{C_+} \bar{z}^{n+1} \bar{T}(\bar{z}) \,\mathrm{d}{\bar{z}}\notag\\
&=\frac{1}{2 \pi i} \int_{C} z^{n+1} \mathbb{T}(z) \,\mathrm{d}{z}\ ,
\end{align}
the integration is performed only on the upper half semicircle $C_+$ and $L_n^{(H)}$ should satisfy the Virasoro algebra. Note that in BCFT, we only have one copy of Virasoro algebra. Under the conformal mapping, the Hamiltonian transforms as
\begin{align}
    H_{\alpha\beta}=\left(\frac{\pi}{L}\right)\frac{1}{2\pi i}\int_C \frac{\mathrm{d}z}{z}(z^2\mathbb{T}(z)-\frac{c}{24})=\frac{\pi}{L}(L^\text{(H)}_0-\frac{c}{24}).
\end{align}
The eigenstates of the Hamiltonian are then those created by acting boundary operators on the vacuum
\begin{align}
    \left|\Psi^{\alpha\beta}_k\right\rangle&=\lim _{x \rightarrow 0} \Psi_k^{\alpha\beta}(x)\left|0\right\rangle\notag\\
     \left\langle\Psi^{\alpha\beta}_k\right|&=\lim _{x \rightarrow \infty} x^{2 h_k}\left\langle 0\right| \Psi^{\beta\alpha}_k(x)
\end{align}
where $\Psi^{\alpha\beta}_k$ is the boundary condition changing operator (or boundary operator in short) inserted on the real axis with $\beta$ b.c. on the left-hand side and $\alpha$ b.c. on the right-hand side. The lower label $k$ denotes how it transforms under the Virasoro algebra.
 The vacuum $\left| 0\right\rangle$ is generally not an $SL(2)$ invariant vacuum and should be labeled by boundary conditions.
However, since we will not merge different pairs of boundary conditions by the Hamiltonian, the omission of this label should not be ambiguous. 
The energy of each state is determined by its conformal dimension $h_k$:
\begin{align}
H_{\alpha\beta}\left|\Psi^{\alpha\beta}_k\right\rangle=\frac{\pi}{L}(h_k-\frac{c}{24})\left|\Psi^{\alpha\beta}_k\right\rangle\ .
\end{align}

\subsection{Finite-size corrections from perturbation theory}\label{Sec_PBCFT}
To calculate the finite-size corrections, like the PBC case \eqref{eq_HPBC}, we consider an effective Hamiltonian near the RG fixed point:
\begin{align}
    \mathcal{H}_{\alpha \beta}^{\text {eff }}(L)=H_{\alpha \beta}^{\mathrm{CFT}}+\sum_i g_i \int_0^L \mathrm{d} v\, \Phi_i(v)+\sum_j g_j^L \Psi_j^{\alpha \alpha}(0)+\sum_j g_j^R \Psi_j^{\beta \beta}(L),
\end{align}
where $\Psi_j^{\alpha\alpha}$ and $\Psi_j^{\beta\beta}$ are boundary operators living on $\alpha$ and $\beta$ boundaries, respectively.
Performing the conformal mapping, we have
\begin{align}\label{H_eff}
H^{\rm eff}_{\alpha\beta}(L)&=\frac{\pi}{L}(L_0-\frac{c}{24})+\sum_i g_i\left(\frac{L}{\pi}\right)^{1-\Delta_{i}} \int_0^\pi \mathrm{d}{\theta}\, e^{-i\theta s_{i}}\Phi_{i}(\exp (i \theta)) \nonumber\\
&\quad+\sum_i g^L_j\left(\frac{L}{\pi}\right)^{-h_i} \Psi_j^{\alpha\alpha}(1)+\sum_i g^R_j\left(\frac{L}{\pi}\right)^{-h_j} \Psi_j^{\beta\beta}(-1)\ .
\end{align}
This allows us to calculate the finite-size corrections by perturbation theory. For example, for a boundary perturbation located on the left-hand side, the first-order perturbation to the $n$-th excited state is given by
\begin{align}\label{eq_bo_1st}
    \delta\left( \calE_n\right)^\text{(1)}_j &= g^{L}_j \left(\frac{\pi}{L}\right)^{h_j}
    \big\langle\Psi_n^{\alpha\beta}\big|\Psi^{\alpha\alpha}_j(1)\big|\Psi_n^{\alpha\beta}\big\rangle
    =g^{L}_j\left(\frac{\pi}{L}\right)^{h_j}C^{\beta\alpha\alpha}_{njn}\ .
\end{align}
For decedents, boundary $T$, for instance, will have a first-order perturbation as:
\begin{align}
        \delta\left(\calE_n\right)_T^\text{(1)}&= \left(\frac{\pi}{L}\right)^{2}
    \big\langle\Psi_n^{\alpha\beta}\big|g_T^L\left(\mathbb{T}(1)-\frac{c}{24}\right)+g_T^R\left(\mathbb{T}(-1)-\frac{c}{24}\right)\big|\Psi_n^{\alpha\beta}\big\rangle\notag\\
    &=(g_T^L+g_T^R)\left(\frac{\pi}{L}\right)^{2}\left(h_n-\frac{c}{24}\right)
    \end{align}
And the shift of the energy level in the first-order bulk perturbation is given as
\begin{align}\label{bu_1st}
     \delta \left( \calE_n \right)^\text{(1)}_{i}&=g_i\left(\frac{\pi}{L}\right)^{\Delta_i-1} \int_0^\pi \mathrm{d} \theta\, e^{-i\theta s_{i}}\left\langle\Psi_n^{\alpha\beta}\middle|\Phi_{i}(\mathrm{exp}(i\theta))\middle| \Psi_n^{\alpha\beta}\right\rangle\ .
\end{align}
Using the OPE coefficients, Eq.~\eqref{bu_1st} can also be expressed as
\begin{align}
  \delta \left( \calE_n \right)_{i}^{(1)}
     &=g_i\left(\frac{\pi}{L}\right)^{\Delta_i -1}\left\{\int_0^{\pi/2} \mathrm{d} \theta\, e^{i\Delta_i\theta}\sum_p {^\alpha B_{i}^p}C^{\beta\alpha\alpha}_{npn}F^{i,nn}_{p}(1-\exp(2i\theta ))\right.\notag\\
     &\quad\quad\quad\quad\quad\quad+\left. \int_{\pi/2}^{\pi} \mathrm{d} \theta\, e^{i\Delta_i\theta}\sum_p {^\beta B_{i}^p}C^{\beta\beta\alpha}_{npn}F^{i,nn}_{p}(1-\exp(2i\theta ))\right\}\ .
\end{align}

Note that we split the integral in Eq.~\eqref{bu_1st} since we have different boundary conditions located on the real axis.
Unlike the PBC case, regardless of the boundary condition and the state we are considering, Eq.~\eqref{bu_1st} always survives since the identity channel $p=0$ (identity operator $\mathbb{1}$)
contributes non-trivially.

\subsection{Finite-size corrections from bulk \texorpdfstring{$T$}{TT}}
\label{TTbar}

As a concrete example, let us consider the first-order perturbation from the bulk perturbation $T^2$ and $\bar{T}^2$.
Note that they have conformal spin $\pm 4$ and thus are not Lorentz invariant.
Nevertheless, they can appear in the effective theory of a lattice model which does not have the Lorentz invariance at the microscopic level. Here, we only consider the combination $T^2+\bar{T}^2$ since $T^2-\bar{T}^2$ does not respect the spatial inversion (parity) symmetry $E(p)=E(-p)$ \cite{reinickeAnalyticalNonanalyticalCorrections1987}.
Such a term, however, is RG-irrelevant, thereby properly recovering the Lorentz invariance of the system in the infra-red regimes as expected.

Under the PBC, such
perturbation is known to result in the Leading Order (LO) correction to the finite-size scaling (FSS) in the TF-Ising model and the Next-to-Leading Order (NLO)
corrections in the three-state Potts model.
We consider the same perturbation, but this time on a strip geometry:
\begin{align}\label{T}
H_{T^2}&=g\int_0^L(T^2(w)+\bar{T}^2(\bar{w}))\mathrm{d}{y}\notag\\
&=2\pi g\left(\frac{\pi}{L}\right)^3\left[\frac{1}{2\pi i}\int_{C^+}\frac{\mathrm{d}z}{z}\int_z \frac{\mathrm{d}z^\prime(z^2T(z)-\frac{c}{24})({z^\prime}^2T(z^\prime)-\frac{c}{24})}{2\pi i(z^\prime-z)}\right.\notag\\
&\quad\quad\quad\quad\ \ \ \ -\left.\frac{1}{2\pi i}\int_{C^+}\frac{\mathrm{d}\bar{z}}{\bar{z}}\int_{\bar{z}}\frac{\mathrm{d}\bar{z}^\prime(\bar{z}^2\bar{T}(\bar{z})-\frac{c}{24})({\bar{z}^{\prime 2}} \bar{T}(\bar{z}^\prime)-\frac{c}{24})}{2\pi i(\bar{z}^\prime-\bar{z})}\right]\notag\\
&=2\pi g\left(\frac{\pi}{L}\right)^3\left[\frac{1}{2\pi i}\int_{C}\frac{\mathrm{d}z}{z}\int_z \frac{\mathrm{d}z^\prime(z^2\mathbb{T}(z)-\frac{c}{24})({z^\prime}^2\mathbb{T}(z^\prime)-\frac{c}{24})}{2\pi i(z^\prime-z)}\right]\notag\\
&=2\pi g\left(\frac{\pi}{L}\right)^3\left[2 \sum_{n=1}^{\infty} L^\text{(H)}_{-n} L^\text{(H)}_n+L_0^\text{(H)2}-\frac{c+2}{12} L_0^{(H)}+\frac{c(22+5 c)}{2880}\right]\ ,
\end{align}
where we performed the conformal mapping as in  Eq.~\eqref{H_eff} to obtain the 2nd line, and the mode expansion of $\mathbb{T}$ gave the last line \cite{poghosyanShapingLatticeIrrelevant2019}.
Eq.~\eqref{T} is diagonal with respect to the eigenstates of the Virasoro operator $L^{(H)}_0$.
Furthermore, the eigenvalues (diagonal matrix elements) are determined by the conformal dimension $h$
and level of descendant of each Virasoro eigenstate, as given in Eq.~\eqref{T_mat0}.

We find that
the result is similar to the PBC case \cite{reinickeAnalyticalNonanalyticalCorrections1987};
 the result for the OBC is obtained simply by substituting all the $\frac{2\pi}{L}$by $\frac{\pi}{L}$
and the sum of two copies of Virasoro algebra by the only existing one on the upper half-plane.
This is due to the conformal invariance of the boundary
\begin{align}
    T(z) = \bar{T}(\bar{z}),
\end{align}
when the boundary is the real axis.
Thanks to this property, $\bar{T}(z)$ on the upper half plane can be identified with the analytic continuation of $T(z)$ to the fictitious lower half-plane, which is often referred to as the mirror image method.
The calculations using the full complex plane in
Ref.~\cite{izmailianFinitesizeCorrectionsIsing2010,izmailianUniversalAmplitudeRatios2012,izmailianBoundaryConditionsAmplitude2009}
can be interpreted as the mirror image method, leading to the correct ratio between the FSC of different energy levels.
In fact, for each operator $f[T](z)$ belonging to identity family, \eqref{T} can be generalized easily for perturbations like $f[T]+f[\bar{T}]$
\begin{align}
    H_{f{[T]}}^\text{OBC}&=g_\text{OBC}\int_0^L(f[T](w)+f[\bar{T}](\bar{w}))\,\mathrm{d}{y}\notag\\
    &=2g_\text{OBC}L\left\{\frac{1}{2\pi i}\int_{C^+}f\left[\left(\frac{\pi}{L}\right)^2\left(z^2T(z)-\frac{c}{24}\right)\right]\frac{\mathrm{d}z}{z}\right.\notag\\
    &\ \quad\quad\quad\quad\left.-\frac{1}{2\pi i}\int_{C^+}f\left[\left(\frac{\pi}{L}\right)^2\left(\bar{z}^2T(\bar{z})-\frac{c}{24}\right)\right]\frac{\mathrm{d}\bar{z}}{\bar{z}}\right\}\notag\\
    &=2g_\text{OBC}L\left\{\frac{1}{2\pi i}\int_{C}f\left[\left(\frac{\pi}{L}\right)^2\left(z^2\mathbb{T}(z)-\frac{c}{24}\right)\right]\frac{\mathrm{d}z}{z}\right\}\notag\\
    &=:W_f[2L,\{L^\text{(H)}_n\},g_\text{OBC}]\ ,
\end{align}
where $W_f$ is determined by $f$. On the other hand, in the PBC case,
\begin{align}
    H_{f{[T]}}^\text{PBC}&=g_\text{PBC}\int_0^L(f[T](w)+f[\bar{T}](\bar{w}))\,\mathrm{d}{y}\notag\\
    &=g_\text{PBC}L\left\{\frac{1}{2\pi i}\int_{C}f\left[\left(\frac{2\pi}{L}\right)^2\left(z^2T(z)-\frac{c}{24}\right)\right]\frac{\mathrm{d}z}{z}\right.\notag\\
    &\ \ \ \quad\quad\quad\left.+\frac{1}{2\pi i}\int_{C}f\left[\left(\frac{2\pi}{L}\right)^2\left(\bar{z}^2T(\bar{z})-\frac{c}{24}\right)\right]\frac{\mathrm{d}\bar{z}}{\bar{z}}\right\}\notag\\
    &=W_f[L,\{L_n\},g_\text{PBC}]+W_f[L,\{\bar{L}_n\},g_\text{PBC}]\ .
\end{align}

The Lorentz invariance-breaking marginal bulk perturbation $T + \bar{T}$ is also allowed in lattice models. With the help of the above results, it just gives rise to an renormalization of the spin-wave velocity (``speed of light''), like the PBC case.

One can also consider the $T\bar{T}$ perturbation. This time, the image method fails to apply, but one can still perform the mode expansion to obtain
\begin{align}
    H_{T\bar{T}}&=g_{T\bar{T}}\int_0^LT(w)\bar{T}(\bar{w})\,\mathrm{d}y\notag\\
    &=2\pi g_{T\bar{T}}\left[\frac{1}{2\pi i}\int_{C^+}\left(\frac{\pi}{L}\right)^3\left(z^2\mathbb{T}(z)-\frac{c}{24}\right)\left(z^{*2}\mathbb{T}(z^*)-\frac{c}{24}\right)\frac{\mathrm{d}z}{z}\right]\notag\\
    &=\pi g_{T\bar{T}}\left(\frac{\pi}{L}\right)^3\left[\sum_{n\in \mathbb{Z}}L^\text{(H)}_n L_n^\text{(H)}-\frac{c}{12}L^\text{(H)}_0+\left(\frac{c}{24}\right)^2\right]
\end{align}
In evaluating the diagonal matrix element, the only non-vanishing term in the summation is $L^\text{(H)}_0 L_0^\text{(H)}$. Overall, we have
\begin{align}\label{TT_mat}
 \delta \left( \calE_n \right)_{T\bar{T}}^\text{(1)}=\left\langle \Delta,r \middle|H_{T\bar{T}}\middle|\Delta,r\right\rangle=\pi g_{T\bar{T}}\left(\frac{\pi}{L}\right)^3\left(\Delta+r-\frac{c}{24}\right)^2\ .
\end{align}
Such FSC, if it presents, is at the same order as the 1st order perturbation from $T^2+\bar{T}^2$.
\subsection{UV divergence and regularization}
\label{Sec_uv}
We close this section by considering some corrections that need regularisation of UV divergence.
There are two kinds of UV-divergence in general.
One comes from the higher-order perturbation, where we consider the contribution of two identical operators approaching each other in the temporal direction.
As the easiest example, let us calculate the second-order perturbation to the ground-state energy from a pair of boundary operators $V =g^L_\Psi \Psi(0)+g^R_\Psi \Psi(L)$. For simplicity, we assume that
$\Psi$ is a primary operator with conformal dimension $h_\Psi > 0$, but the result should be easily generalized to conformal descendants. 

For any excited state $n$, if this state is non-degenerated in the absence of the perturbation, we have
\begin{align}\label{bo_2nd}
    \delta \left(\calE_n^{\alpha\beta}\right)_\Psi^{(2)} &=
    -\sum_{i\neq n}\frac{\langle n|V| i\rangle\langle i|V| n\rangle}{E_i^{(0)}-E_n^{(0)}}\notag\\
    &=-\int_0^\infty \mathrm{d}\tau \left\{\sum_{i=0}^\infty \left.\exp [-(E_i^{(0)}-E_n^{(0)})\tau]\right.\right.
    \langle n|V| i\rangle\langle i|V| n\rangle\notag\\
    &\quad\left.-\:\langle n|V| n\rangle^2-2\sum_{i<n}\cosh[(E_n^{(0)}-E_i^{(0)})\tau]\langle n|V| i\rangle\langle i|V| n\rangle\right\}\ .
\end{align}
We will only consider the case where $\alpha=\beta$. Then, for the ground state, which corresponds to the identity operator, the second term in \eqref{bo_2nd} vanishes by the fusion rule.
The third term does not exist since $i\geq 0$
, so the second-order perturbation can be obtained simply from a 2-point function,
\begin{align}
    \delta \left( \calE^{\alpha\alpha}_0 \right)_\Psi^{(2)} &=-\int_0^\infty\mathrm{d}\tau\langle 0|V(0)V(\tau)|0\rangle\notag\\
    \label{eq_bo_2nd}&=- \left(\frac{\pi}{L}\right)^{2h_\Psi}\int_0^\infty\mathrm{d}\tau
    \left\{\frac{{g^L_\Psi}^2+{g^R_\Psi}^2}{\left[2\sinh\left(\frac{\pi \tau}{2L}\right)\right]^{2h_\Psi}}+\frac{2g^L_\Psi g^R_\Psi}{\left[2\cosh(\frac{\pi\tau}{2L})\right]^{2h_\Psi}}\right\}.
\end{align}
The first term is singular when $h_\Psi\geq\frac{1}{2}$. To regularise it, we introduce a UV-cutoff $b$, which can be identified as the discreteness nature of the lattice model, so $b\sim a$
\begin{align}\label{fsc_gs_bound_1_reg}
    \int_b^\infty\frac{\mathrm{d}\tau}{\left[2\sinh(\frac{\pi \tau}{2L})\right]^{2h_\Psi}}=\left(\frac{L}{\pi}\right)\left\{\frac{\Gamma(1/2-h_\Psi)\Gamma(h_\Psi)}{2^{2h_\Psi}\sqrt{\pi}}+\left(\frac{L}{\pi b}\right)^{2h_\Psi-1}\left[\frac{\Gamma(1/2-h_\Psi)}{2\Gamma(3/2-h_\Psi)}+O\left(\frac{b}{L}\right)\right]\right\}\ .
\end{align}
We keep only the first term as a regularisation since it is independent of the UV cutoff. We note that the only non-vanishing singularity in the thermodynamic limit $\frac{b}{L}\ll1$ comes from the second term and can be organized into the boundary energy. When $h_\Psi=\frac{1}{2}$, we have log correction:
\begin{align}\label{fsc_gs_bound_2_reg}
    \int_a^\infty\frac{\mathrm{d}\tau}{2\sinh(\frac{\pi \tau}{2L})}=\left(\frac{L}{\pi}\right)\left[2\log2-\log\left(\frac{\pi b}{L}\right)+O\left(\frac{b}{L}\right)\right].
\end{align}
On the other hand, integration for the second term in \eqref{eq_bo_2nd} can be carried out
directly for $h_\Psi>0$:
\begin{align}
    \int_0^\infty\frac{\mathrm{d}\tau}{\left[2\cosh\left(\frac{\pi \tau}{2L}\right)\right]^{2h_\Psi}}=\left(\frac{L}{\pi h_\Psi}\right){}_2F_1(h_\Psi;2h_\Psi;1+h_\Psi;-1)\ .
\end{align}
Overall, the regularised FSC is
\begin{align}
    \delta \left( \calE_0^{\alpha\alpha}\right)_\Psi^{(2)} \sim - \left(\frac{\pi}{L}\right)^{2h_\Psi-1}\left[\left({g^L_\Psi}^2+{g^R_\Psi}^2\right)
    \left(\frac{\Gamma(1/2-h_\Psi)\Gamma(h_\Psi)}{2^{2h_\Psi}\sqrt{\pi}} + \text{const.} \right)\right.\notag\\
    +\left.
    \frac{2g^R_\Psi g^L_\Psi}{h_\Psi}{}_2F_1(h_\Psi;2h_\Psi;1+h_\Psi;-1)\right]
\end{align}
for $h_\Psi>\frac{1}{2}$, and
\begin{align}\label{h=1/2}
    \delta \left( \calE_0^{\alpha\alpha} \right)^{(2)}_\Psi =
    - \left(\frac{\pi}{L}\right)^{2h_\Psi-1}\left[\left({g^L_\Psi}^2+{g^R_\Psi}^2\right)
    \left(\text{const.}+ \log{L}\right)+g^L_\Psi g^R_\Psi \pi\right]
\end{align}
for $h_\Psi =\frac{1}{2}$.
The UV divergent terms are independent of $L$ and can be absorbed as non-universal boundary
energies.
The finite constant $2\log{2}$ in Eq.~\eqref{fsc_gs_bound_2_reg} is additive to the
logarithmic divergent term $\log{a}$ which is non-universal.
Thus there is no universal constant that is predictable from field theory.
On the other hand, $\log{L}$ term gives a universal FSC.
This particular pattern turns out to be crucial in describing the boundary RG flow of the TF-Ising model
with boundary fields discussed in Sec.~\ref{sec:sum_ising}.

The other kind of UV divergence comes from the spatial integral of a bulk operator,
when the bulk operator $\Phi$ is diagonal, namely if the holomorphic and antiholomorphic
parts of $\Phi$ are the same operators.
In this case, the one-point function of $\Phi$ near the boundary generally diverges as
\begin{align}
    \left\langle\Phi(z)\right\rangle_\alpha\sim\frac{1}{\text{Im}{z}^{\Delta_\Phi}}\ ,
\end{align}
where $\Delta_\Phi$ is the total scaling dimension of $\Phi$,
since it can be mapped to a 2-point function of holomorphic operators across the boundary.
This can be also understood as the contribution of the identity channel
$\Psi^{\alpha\alpha}_0 = \mathbb{1}$ in the expansion~\eqref{bb-OPE}.

For simplicity, we will consider its first-order perturbation to the ground-state energy with identical boundary condition $\alpha$ on each side. For a bulk field $\Phi_i$ transforms as $i \bigotimes \bar{i}$, since $C^{\alpha\alpha\alpha}_{0p0}=\delta_{0,p}$, the only contributing channel in \eqref{bu_1st} is the identity channel:
\begin{align}
    \delta \left( \calE_0^{\alpha\alpha} \right)^{(1)}_\Phi&=g \left(\frac{\pi}{L}\right)^{\Delta_\Phi-1} \int_0^\pi \mathrm{d}{\theta}
    \left\langle \Phi_i(\mathrm{exp}(i\theta))\right\rangle_\alpha\notag\\
    \label{bulk1}&=g \left(\frac{\pi}{L}\right)^{\Delta_\Phi-1}{}^{\alpha}B_{i}^{\mathds{1}}\int_0^\pi \frac{\mathrm{d}\theta}{(2\sin\theta)^{\Delta_\Phi}}\ .
\end{align}
This time, by introducing the same UV cutoff, we have
\begin{align}\label{fsc_gs_bulk_1_reg}
    \int_{\frac{\pi b}{L}}^{\pi-\frac{\pi b}{L}} \frac{\mathrm{d}\theta}{(2\sin\theta)^{\Delta_\Phi}}=\frac{\sqrt{\pi}\Gamma(1/2-\Delta_\Phi/2)}{2^{\Delta_\Phi}\Gamma(1-\Delta_\Phi/2)}-\left(\frac{L}{\pi b}\right)^{\Delta_\Phi-1}\left[\frac{\Gamma(1/2-\Delta_\Phi/2)}{2^{\Delta_\Phi}\Gamma(3/2-\Delta_\Phi/2)}+O(b^2/L^2)\right]\ ,
\end{align}
where the first term is kept as a regularised value. We note that the second term can be again organized into the boundary energy, like \eqref{fsc_gs_bound_1_reg} and \eqref{fsc_gs_bound_2_reg}. So the regularised FSC is
\begin{align}
    \delta \left( \calE_0^{\alpha\alpha} \right)^{(1)}_\Phi &=g \left(\frac{\pi}{L}\right)^{\Delta_\Phi-1}{}^{\alpha}B_{i}^{\mathds{1}}\frac{\sqrt{\pi}\Gamma(1/2-\Delta_\Phi/2)}{2^{\Delta_\Phi}\Gamma(1-\Delta_\Phi/2)}\ .
\end{align}

\section{Local shift of the system size and boundary energy}\label{sec_L}

In this section, we examine two crucial ingredients in relating the spectrum of the lattice model to the effective field theory: the effective system size and boundary energy.

As we have reviewed in Sec.~\ref{sec_perturbation}, we can derive the finite-size scaling including subleading corrections, based on the (B)CFT defined on a segment of
a finite length $L$. Naively, the length $L$ is identified with $Na$ in a lattice
mode realization, where $N$ is the
number of sites in the lattice model, and $a$ is the lattice spacing.
However, as we will demonstrate below, there is generally an ambiguity of the order of
the lattice spacing $a$ in the identification; we need to use the appropriate
effective system size $L$ for each concrete lattice realization.

Our ansatz is similar to the concept called extrapolation length $\tau_0$ which measures the deviation of the actual boundary state $\left|\psi\right\rangle\propto e^{-\tau_0 H}\left|B\right\rangle$ from the boundary RG fixed point 
 $\left|B\right\rangle$ \cite{cardyQuantumQuenchesCritical2016} and describes how the field extrapolates to a boundary $\tau_o$ lattice spacing beyond the edges of the system \cite{ziffEffectiveBoundaryExtrapolation1996}. In the study of FSC, a similar method was carried out in Ref.~\cite{campostriniFinitesizeScalingQuantum2014,campostriniQuantumIsingChains2015} to formally remove the $O(1/N^2)$ order FSC in the TF-Ising model.

 This paper 
generalizes this concept further to all the 1+1d quantum spin chains at their critical point with all conformally invariant boundary conditions. In Sec.~\Ref{Sec_L_A}, we introduce our ansatz as an ambiguity of defining the system size in the continuum limit and briefly discuss its significance when generalizing Cardy's perturbation theory to the OBC case. Sec.~\ref{Sec_E_A} is dedicated to reinterpreting the boundary energy in this new framework.
 
\subsection{Local shift of the system size \texorpdfstring{$\delta L$}{lalphabeta}}\label{Sec_L_A}
For an $N$-site quantum chain with open boundary conditions $\alpha$ and $\beta$, the energy spectrum take the form of
\begin{align}\label{En_lattcie}
E^{\alpha\beta}_n(N)=\varepsilon_\alpha+\varepsilon_\beta+\varepsilon_0 N+\frac{\pi v}{N}(x_n-\frac{c}{24})+\sum_i a_n^{i} N^{-b_n^i}\ ,
\end{align}
where $\varepsilon_0$ is the energy per site and $\varepsilon_{\alpha/\beta}$ is the boundary energy. The 
remaining terms are attributed to the perturbations in the effective Hamiltonian \eqref{H_eff}. To relate Eq.~\eqref{H_eff} and Eq.~\eqref{En_lattcie}, it seems tempting to set
\begin{align}
    L=Na
\end{align}
as in the PBC case. However, it is noticed that in the presence of the open boundaries, there is $O(1/N^2)$ FSC in the TF-Ising model, and it can be absorbed into the fourth term in the RHS of \eqref{En_lattcie} formally by simply redefining an effective system size $L_\text{eff}\neq Na$ as motivated in Ref.~\cite{campostriniFinitesizeScalingQuantum2014,campostriniQuantumIsingChains2015}.

The physical motivation for such a redefinition can be interpreted as follows. Consider an $N$-site quantum chain with PBC; the length of the system is naturally defined as the lattice spacing $a$ multiplied by $N$
\begin{align}\label{L_PBC}
    L= Na\ .
\end{align}
In the continuum limit, since we do not need to attach the end of the space to anything, the system size $L$ should also follow Eq.~\eqref{L_PBC}. 
On the other hand, to define the continuum limit of an open chain, there seems to be an ambiguity:
\begin{align}\label{L_OBC}
    L\overset{?}{\sim} (N-1)a\ .
\end{align}
Since we need to attach the space to some boundary conditions. This time, we might have the following form of the system size in the continuum limit
\begin{align}\label{defl}
    L\sim (N+(\delta L)_\alpha+(\delta L)_\beta)a\ ,
\end{align}
    where $L_\alpha$ and $L_\beta$ are "local shifts of the system size" that are determined only by the specific boundary conditions $\alpha$ and $\beta$ applied on the ends of the spin chain, respectively. Apparently, $\delta L\sim O(1)$. In this way, we claim that to match the lattice and BCFT spectrum, one needs to fix the system size in BCFT to be
\begin{align}\label{def_L}
    L=\left[N+(\delta L)_\alpha+(\delta L)_\beta\right]a\ .
\end{align}
So the energy spectrum \eqref{En_lattcie} can be reorganized into
\begin{align}\label{En_redef}
    E_n^{\alpha\beta}(N)&=\frakE_\alpha+\frakE_\beta+\varepsilon_0(N+(\delta L)_\alpha+(\delta L)_\beta)+\frac{\pi v}{N+(\delta L)_\alpha+(\delta L)_\beta}(x_n-\frac{c}{24})\notag\\
    &+\sum_i a_n^{\prime i} \left[N+(\delta L)_\alpha+(\delta L)_\beta\right]^{-b_n^{\prime i}}\ ,
\end{align}
where $\frakE_{\alpha/\beta}\coloneqq\varepsilon_{\alpha/\beta}-\varepsilon_0 (\delta L)_{\alpha/\beta}$ are called the redefined boundary energy and will be discussed further in the next subsection. $\{a^{\prime i}_n\}$ and $\{b^{\prime i}_n\}$ are the coefficients and powers in expanding $E_n^{\alpha\beta}(N)$ as a series of $N+(\delta L)_\alpha+(\delta L)_\beta$. Here, we remark that, unlike the PBC case, one can only explain these coefficients through the perturbation theory if one chooses the correct $(\delta L)_{\alpha/\beta}$. Otherwise, the set of the coefficients $\{a^{\prime i}_n\}$ and $\{b^{\prime i}_n\}$ will be a mixture of the Taylor expansion of all preceding perturbations. This feature of the systems with open boundaries will make applying Cardy's perturbation theory significantly harder than the PBC case. This ambiguity is non-universal and greatly depends on the concrete realization of the boundary conditions in the lattice model, and there is no general
theoretical description of the shift $\delta L$.
Nevertheless, as we have seen, sometimes the shifts take simple
values and/or obey a certain relation when an exact solution
or a Kramers-Wannier type duality is applicable to the lattice model.

\subsection{Redefinition of boundary energy \texorpdfstring{$\frakE$}{e}}\label{Sec_E_A}
Another consequence of renormalizing the system as in~\eqref{En_redef} is that part of the boundary energy $\varepsilon_{\alpha/\beta}$ is merged into the bulk energy as
\begin{align}
    \varepsilon_{\alpha/\beta}=\frakE_{\alpha/\beta}+\varepsilon_0 (\delta L)_{\alpha/\beta}\ .
\end{align}
The remaining part $\frakE_{\alpha/\beta}$ is the redefined boundary energy.

As we remarked in Sec.~\ref{Sec_uv}, in the presence of boundaries,
the bulk and boundary irrelevant perturbations give rise to 
UV divergent corrections which are independent of the system size $L$.
We can interpret such diverging constant corrections as renormalizations
of the non-universal boundary energy.

The discussion in Sec.~\ref{Sec_uv} is extended to general $n$-th order
perturbation theory as follows.
The $n$-th order perturbation is generally given by a $2n-1$-fold integral
of the $n$-point correlation function, which scales as $\text{(length)}^{-n \Delta}$.
Although naively the system size $L$ is the only length scale of the problem,
we also need to introduce the UV cutoff $b$, which is of the order of the lattice
spacing $a$ for a lattice model to regulate UV divergence.

The dimensional analysis and the physical argument
suggest that a finite-size energy eigenvalue 
will have the $n$-th order corrections due to the bulk perturbation $g$ as
\begin{align}
    \delta \left( \calE \right)^{(n)}_g \sim
    \text{const.} \frac{1}{L} \left( \frac{g}{L^{\Delta-2}} \right)^n
    + \text{const.} \frac{1}{b} \left( \frac{g}{b^{\Delta-2}} \right)^n
    + \text{const.} \frac{L}{b^2} \left( \frac{g}{b^{\Delta-2}} \right)^n,
\end{align}
where $\Delta$ is the total scaling dimension of the bulk perturbation.
The first term is the cutoff-independent universal correction to the energy level.
The term proportional to $L$ leads to the renormalization of the non-universal
bulk energy density.
On the other hand, the constant ($L$-independent) term corresponds to
the renormalization of the non-universal boundary energy.
We note that, under the periodic boundary condition, the constant term is absent
as there is no boundary energy.

Similarly, for a boundary perturbation $g^{L/R}=g^B$, the $n$-th order corrections take
the form
\begin{align}
    \delta \left( \calE \right)^{(n)}_{g^B} \sim
        \text{const.} \frac{1}{L} \left( \frac{g^B}{L^{h-1}} \right)^n
    + \text{const.} \frac{1}{b} \left( \frac{g^B}{b^{h-1}} \right)^n ,
\end{align}
where $h$ is the scaling dimension of the boundary perturbation.
Again, in addition to the universal, cutoff-independent correction to the scaling,
there is a divergent $L$-independent contribution which can be absorbed by
the renormalization of the non-universal boundary energy.

As we have observed, the analysis of the energy spectrum becomes
more complicated for open boundary conditions compared to that for the periodic 
boundary conditions, owing to the presence of both bulk and boundary perturbations
and the non-universal boundary energy.
The non-universal boundary energy can be however removed by considering the
energy differences.
On the other hand, the effective system size must be appropriately chosen
in order to minimize the corrections even when considering the energy differences.

\section{Discussion and future directions}
\label{sec_dis}
In this paper, we first systematically investigated finite-size corrections to the spectrum of critical one-dimensional quantum systems in terms of BCFT with
bulk and boundary perturbations.
The ubiquitous $O(1/N^2)$ FSC is attributed to the energy-momentum tensor perturbation
at the boundary, which can be eliminated by a shift of the effective system size.
We have compared our theoretical results with the exact solution and numerical data
for the critical TF-Ising and three-state Potts chains
and found very good agreements.
We confirmed that the bare coupling constant of the bulk irrelevant perturbation remains the same regardless of the presence of the open boundary.
In the three-state Potts model, we found the first existing irrelevant boundary perturbation $X$ and evaluated its effect on the spectrum, together with the already known bulk perturbation $X\bar{X}$.

When one adds a relevant boundary perturbation, the system might flow into another BCFT fixed point with another conformally invariant boundary condition, under the RG flow. This is called the boundary RG flow~\cite{grahamRenormalisationGroupFlows2000}. It can be achieved on the lattice by applying a longitudinal magnetic field on the boundary for the Ising and three-state Potts model.
In fact, we confirmed the agreement between our perturbed BCFT result and
the exact solution for the critical Ising chain with
boundary longitudinal magnetic fields, in the leading order of the boundary field.
It would be interesting to extend the analysis of the boundary RG flow to
higher orders, and also to non-perturbative level.
This would be rather challenging since the non-universal quantities
such as the shift of the effective length $\delta L$ and the boundary energy
$\frakE$ will also be renormalized in the boundary RG flow.
We hope to develop a new formulation to study the entire region of the
boundary RG flow in the future, handling the non-universal quantities appropriately.

\section*{Acknowledgements}
We thank Natalia Chepiga and Yuan Miao for their helpful discussions.


\paragraph{Funding information}
Y.~L. is supported by the Global Science Graduate Course (GSGC) program of the University of Tokyo.
A.~U.~is supported by the MERIT-WINGS Program at the University of Tokyo, the JSPS fellowship (DC1). He was supported in part by MEXT/JSPS KAKENHI Grants No.\ JP21J2052.
The work of M.~O. is supported by JSPS KAKENHI Grant Nos. JP23H01094, JP23K25791 and JP24H00946. 
This research was supported in part by grant NSF PHY-2309135 to the Kavli Institute for Theoretical Physics (KITP), during the visit there by M.~O. and Y.~L. 
A part of the computation in this paper has been done using the facilities of the Supercomputer Center, the Institute for Solid State Physics, the University of Tokyo.

\begin{appendix}
\section{Kramers-Wanier duality in TF-Ising model}\label{app:K-W}
Kramers-Wannier transformation can be implemented
as a unitary transformation on an open chain~\cite{K-W,Linhao-MO-Yunqin}.
It maps the spin operators as
\begin{align}
\sigma^x_j &=
     \begin{cases}
     \tilde{\sigma}_{j-1}^z \tilde{\sigma}_j^z, & j=2, \ldots, N\\
     \tilde{\sigma}_1^z, & j=1
     \end{cases}\ ,\notag
\\
\sigma^z_j &= \prod_{k=j}^{N} \tilde{\sigma}_k^x\ .
\end{align}
Applying this to the Ising chain~\eqref{eq:IsingChain} of $N$ sites
with no boundary fields ($\zeta_L=\zeta_R=0$), we obtain the dual Hamiltonian:
\begin{equation}
H = - \sum_{j=1}^{N-1} \tilde{\sigma}^z_j \tilde{\sigma}^z_{j+1}
- \sum_{j=1}^{N-1} \tilde{\sigma}^x_j
    - \tilde{\sigma}^z_1\ .
\end{equation}
At the left end of the chain (site $1$), it has
a boundary longitudinal field $\zeta_L=1$.
On the other hand, at the right end of the chain (site $N$), there is neither
a transverse field nor a longitudinal field.
As a consequence, $\tilde{\sigma}^z_N$ commutes with the Hamiltonian.
We can consider each sector $\tilde{\sigma}^z_N=\pm 1$ separately~\cite{campostriniQuantumIsingChains2015}.
The sector $\tilde{\sigma}^z_N=1$ is equivalent to the Ising chain~\eqref{eq:IsingChain}
of $N-1$ sites with the parallel boundary fields $\zeta_L=\zeta_R=1$, whereas
the sector $\tilde{\sigma}^z_N= -1$ is equivalent to the Ising chain~\eqref{eq:IsingChain}
of $N-1$ sites with the antiparallel boundary fields $\zeta_L=\zeta_R=-1$.

Noticing that having longitudinal field $\zeta=1$ ($\zeta=-1$) is equivalent to having a extra imaginary spin fixed to $\uparrow$ ($\downarrow$) at the end, we obtain the exact identities~\eqref{eq:KW_parallel} and~\eqref{eq:KW_antiparallel}.

\section{Ground-state energy with small boundary magnetic field}\label{app: magnetic}
Following \cite{campostriniQuantumIsingChains2015}, the scaled momentum $b$ satisfies
\begin{align}
    \tan b=\frac{1}{2 b} \frac{\zeta_b{ }^4-b^2}{\zeta_b{ }^2}\ .
\end{align}
At $\zeta=\zeta_b=0$ (free boundary condition), the solutions are
\begin{align}
& b=0\ , \notag\\
& b=\frac{2 n-1}{2} \pi\quad (n=1,2, \ldots)\ .
\end{align}

The ground-state energy is given by the sum of the zero-point energies as
\begin{align}
E_{G S}\left(\zeta_b=0\right) & =-\frac{1}{2} \sum_{n=1}^{\infty} \epsilon\left(b=\frac{2 n-1}{2} \pi\right) \notag\\
& =-\frac{v \pi}{2} \sum_{n=1}^{\infty} \frac{2 n-1}{2 L}\ .
\end{align}
This is evidently divergent, but we can regularize it by introducing the soft cutoff $e^{-\lambda \varepsilon}$
\begin{align}
E_{G S}\left(\zeta_b=0\right) & =-\frac{\pi v}{2} \sum_{n=1}^{\infty} \frac{2 n-1}{2 L} e^{-\lambda v(2 n-1) /(2 L)} \notag\\
& =-\frac{\pi v}{2} \frac{e^{\frac{\lambda v}{2 L}}\left(e^{\lambda v / L}+1\right)}{2 L\left(e^{\lambda v/ L}-1\right)^2} \notag\\
& \sim-\frac{\pi v}{2}\left\{\frac{L}{\lambda^2v^2}+\frac{1}{24 L}-\frac{7 \lambda^2v^2}{1920 L^3}+O\left(\left(\frac{1}{L}\right)^5\right)\right\}\ .
\end{align}
The first term, which is divergent, is the non-universal bulk energy. In this expression, we do not find the non-universal boundary energy because we have taken the continuum limit first. The second term in the expansion is the universal Casimir energy. The universal part reads
\begin{align}
    E_{GS}(\zeta_b-0)\sim-\frac{\pi v}{48L}\ ,
\end{align}
which agrees with the CFT result $-\frac{\pi vc}{24L}$ with the Ising central charge $c=1/2$.

For a finite but small $\zeta_b$, the zero mode is now shifted to
\begin{align}
    b={\zeta_b}^2+O({\zeta_b}^4)\ ,
\end{align}
and the non-zero modes take the form of
\begin{align}
    b=\frac{(2n-1)\pi}{2}+\frac{4}{\pi(2 n-1)} \zeta_b{ }^2+O\left(\zeta_b{ }^4\right) \quad(n=1,2, \ldots)\ .
\end{align}
Therefore, the ground-state energy in the parallel/antiparallel field configurations
are
\begin{align}
E_{G S}\left(\zeta_b\right) & \sim-\frac{1}{2}\left[ \pm \epsilon\left(b=\zeta_b^2\right)+\sum_{n=1}^{\infty} \epsilon\left(b=\frac{(2 n-1) \pi}{2}+\frac{4}{\pi(2 n-1)} \zeta_b^2\right)\right]\notag \\
& =E_{G S}\left(\zeta_b=0\right) \mp \frac{v}{2 L} \zeta_b^2-\frac{v}{2 L} \sum_{n=1}^{\infty} \frac{4}{\pi(2 n-1)} \zeta_b{ }^2\notag \\
& =E_{G S}\left(\zeta_b=0\right) \mp \frac{v}{2} \zeta^2-\frac{2v}{\pi}\zeta^2 \sum_{n=1}^{\infty} \frac{1}{2 n-1}\ ,
\end{align}
where the double sign corresponds to parallel/antiparallel field configuration, respectively. The second correction term to the ground-state energy is a logarithmically divergent series, which can again be regularized with the same exponential soft cutoff:
\begin{align}
\sum_{n=1}^{\infty} \frac{1}{2 n-1} 
&=\sum_{n=1}^{\infty} \lim_{\frac{\lambda}{L}\to0}\frac{\exp \left(-\frac{ \lambda v(2 n-1)}{ 2L}\right)}{2 n-1}\notag  \\
&=\lim_{\frac{\lambda}{L}\to0}\operatorname{ArcTanh}\left(e^{-\frac{\lambda v}{2L}}\right)\notag \\
& \sim \frac{1}{2}\left(\log 2-\log \frac{\lambda v}{ 2L}\right)+\frac{1}{24}\left(\frac{  \lambda v}{ 2L}\right)^2\ .
\end{align}
Thus we obtain Eq.~\eqref{eq:E0_mag_lat}.

\section{Duality in three-state Potts model}\label{app:dual_three}
Following \cite{affleckBoundaryCriticalPhenomena1998}, the dual operators on an open chain are defined as
\begin{align}
    M_j &=
    \begin{cases}
        \tilde{R}^\dagger_{j-1} \tilde{R}_j & j=2, \ldots, N
        \\t
        \tilde{R}^\dagger_1 & j=1 
    \end{cases}\notag\ ,
    \\
    R_j &= \prod_{k=j}^N \tilde{M}_k\ .
\end{align}
They satisfy
\begin{equation}
    \tilde{R}_j \tilde{M}_j = e^{2\pi i/3}  \tilde{M}_j \tilde{R}_j\ ,
\end{equation}
and dual operators on different sites commute as required.

The dual of the free boundary Hamiltonian \eqref{eq:PottsH}
with $\zeta_L=\zeta_R=0$
on an open chain reads
\begin{align}
    \tilde{H} =-\sum_{i=1}^{N-1} \left(  \tilde{M}_i+ \tilde{M}^\dagger_i \right)
    - \sum_{i=1}^{N-1}
    \left( \tilde{R}_i^{\dagger} \tilde{R}_{i+1}+ \tilde{R}_i \tilde{R}_{i+1}^{\dagger} \right)
    - \left( \tilde{R}_1 + \tilde{R}_1^\dagger \right) \ .
\end{align}
Similarly to the Ising case, a longitudinal field $\zeta_L=1$ appears at the left end of the chain while the transverse field is absent at the right end
of the chain.
Therefore, $\tilde{R}_N$ commutes with the dual Hamiltonian and we
can consider the 3 sectors $\tilde{R}_N = 1, e^{\pm 2\pi i/3}$ separately.
As in the case of the Ising model, the longitudinal boundary field $\zeta_L=1$
is equivalent to fixing the end spin to $R=1$ on the chain with an extra site
on the left.
The sector $\tilde{R}_N=1$ gives the lowest energy ground-state and is equivalent to imposing the fixed boundary condition A at both ends.
The sector $\tilde{R}_N=e^{\pm 2\pi i/3}$ is equivalent to imposing the different fixed boundary conditions at each end: AB or BC.

Therefore we find Eqs.~\eqref{eq:3-state_same} and~\eqref{eq:3-state_mixed}.

\section{Some OPE coefficients in boundary \texorpdfstring{$M(6,5)$}{Potts} minimal model}\label{app_OPE}
The OPE coefficients we used in the boundary $M(6,5)$ minimal model are calculated by the method introduced in Ref.~\cite{runkelStructureConstantsDseries2000}, given in terms of the fusion matrices $F$ (see e.g. \cite{Moore1990}) and the modular S matrix (see e.g. \cite{cardyBoundaryConditionsFusion1989}). The fusion matrices are obtained in a recursive way introduced in Ref.~\cite{runkelBoundaryStructureConstants1999}.
We collected those we used in this paper for other possible applications.

\begin{align}
    C^{(3,3)(3,3)(3,3)\ (3,5)_u}_{(3,5)_u(3,1)_u}&=F_{(3,3)(3,5)}\begin{bmatrix}(3,3)&(3,3)\\(3,5)&(3,1)\\ \end{bmatrix}=3(-1+\sqrt{5})\notag\\
    C^{\omega(3,3)(3,3)\ (3,3)_u}_{(3,3)_u(3,1)_u}&=1\notag\\
    C^{(3,3)(3,3)(3,3)\ (3,3)_u}_{(3,3)_u(3,1)_u}&=F_{(3,3)(3,3)}\begin{bmatrix}(3,3)&(3,3)\\(3,3)&(3,1)\\ \end{bmatrix}=\frac{1}{2 (1 - \sqrt{5})}\notag\\
    C^{(1,2)(3,3)(3,3)\ (3,4)_u}_{(3,4)_u(3,1)_u}&=F_{(3,3)(3,4)}\begin{bmatrix}(1,2)&(3,3)\\(3,4)&(3,1)\\ \end{bmatrix}=-3/4\notag\\
    C^{(1,2)(3,1)(3,1)\ (3,4)}_{(3,4)(3,1)}&=F_{(3,5)(3,4)}\begin{bmatrix}(1,2)&(3,5)\\(3,4)&(3,1)\\ \end{bmatrix}=1/8\notag\\
    C^{(1,2)(3,3)(3,3)\ (3,2)_u}_{(3,2)_u(3,1)_u}&=F_{(3,3)(3,2)}\begin{bmatrix}(1,2)&(3,3)\\(3,2)&(3,1)\\ \end{bmatrix}=9/2\notag\\
    C^{(1,2)(3,1)(3,1)\ (3,2)_o}_{(3,2)_o(3,1)_e}&=F_{(3,5)(3,1)}\begin{bmatrix}(1,5)&(3,5)\\(3,1)&(3,1)\\ \end{bmatrix}F_{(3,1)(3,2)}\begin{bmatrix}(1,2)&(3,1)\\(3,2)&(3,1)\\ \end{bmatrix}=-3/4\notag\\
    C^{(1,3)(3,4)(3,4)\ (3,5)_u}_{(3,5)_u(3,1)_u}&=-6C^{\omega(3,1)(3,1)\ (3,5)_e}_{(3,5)_e(3,1)_e}=-6\notag\\
    C^{(1,3)(3,4)(3,4)\ (3,1)_u}_{(3,1)_u(3,1)_u}&=-6C^{\omega(3,1)(3,1)\ (3,1)_o}_{(3,1)_o(3,1)_e}=F_{(3,5)(3,1)}\begin{bmatrix}(1,5)&(3,5)\\(3,1)&(3,1)\\ \end{bmatrix}=9\notag\\
    (A_{(3,1)_e}^{(3,1)(3,1)})^2&=C^{(3,1)(3,1)(3,1)\ (1,1)_e}_{(3,1)_e(3,1)_e}=F_{(3,5)(1,1)}\begin{bmatrix}(3,5)&(3,5)\\(3,1)&(3,1)\\ \end{bmatrix}=\frac{\Gamma\left(-\frac{8}{5}\right) \Gamma\left(\frac{7}{5}\right)}{\Gamma\left(-\frac{7}{5}\right) \Gamma\left(\frac{6}{5}\right)}\notag\\
    (A_{(3,1)_u}^{(3,3)(3,3)})^2&=C^{(3,3)(3,3)(3,3)\ (1,1_u)}_{(3,1)_u(3,1)_u}=F_{(3,3)(1,1)}\begin{bmatrix}(3,3)&(3,3)\\(3,1)&(3,1)\\ \end{bmatrix}=-\frac{9 \Gamma\left(-\frac{3}{5}\right) \Gamma\left(\frac{2}{5}\right)}{\Gamma\left(-\frac{7}{5}\right) \Gamma\left(\frac{6}{5}\right)}\notag\\
    ^{(\omega)} B_{(3,1)\overline{(3,1)}}^1&={}^{(1,2)} B_{(3,1)\overline{(3,1)}}^1=\frac{S_{\varepsilon}}{S_\mathds{1}} = \frac{\sin{(2\pi/5)}}{\sin{(\pi/5)}}\notag\\
    ^{(3,3)} B_{(3,1)\overline{(3,1)}}^1&={}^{(3,2)} B_{(3,1)\overline{(3,1)}}^1=\frac{\sqrt{5}-3}{2} \frac{S_{\varepsilon}}{S_\mathds{1}}\notag\\
    C_{(3,1)\overline{(3,1)}\ (3,1)\overline{(3,1)}}^\mathds{1}&=\frac{S_{\varepsilon}}{S_\mathds{1}}\notag
\end{align}

\end{appendix}



\bibliography{SciPost_Example_BiBTeX_File.bib}

\nolinenumbers

\end{document}